\documentclass[conference]{IEEEtran}

\usepackage{amsmath,amsfonts}
\usepackage{array}
\usepackage[caption=false,font=normalsize,labelfont=sf,textfont=sf]{subfig}
\usepackage{textcomp}
\usepackage{stfloats}
\usepackage{url}
\usepackage{verbatim}
\usepackage{graphicx}
\usepackage{caption}
\usepackage{pgfplots}
\usepackage{tabularx}
\usepackage{tikz}
\usepackage{amsfonts}
\usepackage{multirow}
\usepackage[linesnumbered,ruled,vlined]{algorithm2e}
\usepackage{algpseudocode}
\usepackage{textcomp}
\usepackage{mdframed}
\usepackage{xcolor}
\usepackage{framed}
\usepackage{listings}
\usepackage{graphicx,subcaption}
\usepackage{bbding}
\usepackage{color}
\usepackage{makecell}
\usepackage{enumitem}
\usepackage{minted}
\usepackage{threeparttable}
\usepackage{hyperref}
\usepackage{boxedminipage}
\usepackage{pifont}
\usepackage{tikz}
\usepackage{scalefnt}
\usepackage{booktabs}
\usepackage{makecell}
\usepackage{longtable}
\usepackage[many]{tcolorbox}
\usepackage[utf8]{inputenc}
\usepackage{tikz}
\usepackage{filecontents}
\usepackage{xspace}
\usepackage{listings}
\usepackage{multicol}

\newcommand{\tool}{\emph{FidelityGPT}\xspace}
\newcommand{\degpt}{\emph{DeGPT}\xspace}
\newcommand{\promptz}{\emph{Prompt$_{0}$}\xspace}
\newcommand{\prompte}{\emph{Prompt$_{Eg}$}\xspace}
\newcommand{\promptd}{\emph{Prompt$_{Def}$}\xspace}

\usepackage{balance}
\ifCLASSINFOpdf
\else
\fi
\begin{document}

%
\title{\tool: Correcting Decompilation \\Distortions with Retrieval Augmented Generation}




\author{
\renewcommand{\arraystretch}{1.2}
\begin{tabular*}{\textwidth}{@{\extracolsep{\fill}}cccc@{}}
Zhiping Zhou & Xiaohong Li & Ruitao Feng$^{\ast}$ & Yao Zhang$^{\ast}$ \\
 Tianjin University &  Tianjin University &  Southern Cross University &  Tianjin University \\
 zhou\_zhiping@tju.edu.cn &  xiaohongli@tju.edu.cn &  ruitao.feng@scu.edu.au &  zzyy@tju.edu.cn \\[1em]
Yuekang Li & Wenbu Feng & Yunqian Wang & Yuqing Li \\
 University of New South Wales &  Tianjin University &  Tianjin University &  Tianjin University \\
 yuekang.li@unsw.edu.au &  ianx@tju.edu.cn &  wangyq\_0617@tju.edu.cn &  liyuqing0409@tju.edu.cn
 \end{tabular*}
}

\maketitle

\begingroup
\renewcommand\thefootnote{}%
\footnotetext{%
\hspace*{-\parindent}\rule{\linewidth}{0.4pt}\\
$^{\ast}$ Corresponding authors.\\
\vspace{15ex}
}
\endgroup

\IEEEoverridecommandlockouts
\makeatletter\def\@IEEEpubidpullup{6.5\baselineskip}\makeatother
\IEEEpubid{\parbox{\columnwidth}{
		Network and Distributed System Security (NDSS) Symposium 2026\\
		23 - 27 February 2026 , San Diego, CA, USA\\
		ISBN 979-8-9919276-8-0\\  
		https://dx.doi.org/10.14722/ndss.2026.230989\\
		www.ndss-symposium.org
}
\hspace{\columnsep}\makebox[\columnwidth]{}}

\begin{abstract}
Decompilation is a crucial technique that converts machine code into a human-readable format, facilitating analysis and debugging in the absence of source code. However, this process is hindered by fidelity issues, which can significantly impair the readability and accuracy of the decompiled output. Existing approaches partially addressed these, such as variable renaming and structural simplification, but typically fail to provide adequate detection and correction, especially in complex but practical closed-source binary scenarios. 

To address this, we introduce \tool, a novel framework to improve the accuracy and readability of decompiled code by systematically detecting and correcting discrepancies between decompiled code and its original source. \tool defines distortion prompt templates tailored to closed-source environments and incorporates Retrieval-Augmented Generation (RAG) with a dynamic semantic intensity algorithm. The algorithm identifies distorted lines based on semantic intensity, retrieving similar code from a database. Additionally, a variable dependency algorithm is designed to overcome the limitations of long-context inputs by analyzing redundant variables through their dependencies and integrating redundant variable names into prompt context. These combined techniques establish \tool as the first framework capable of effectively addressing decompilation distortion issues in LLM-based decompilation optimization.
We evaluated \tool on 620 function pairs from a binary similarity benchmark, achieving an average detection accuracy of 89\% and a precision of 83\%. Compared to the current state-of-the-art model, \degpt, which achieved an average Fix Rate (FR) of 83\% and an average Corrected Fix Rate (CFR) of 37\%, \tool demonstrated superior performance. With an average FR of 94\% and an average CFR of 64\%, \tool significantly improves both accuracy and readability, underscoring its effectiveness in enhancing decompilation and its potential to drive advancements in reverse engineering.
\end{abstract}


%
\IEEEpeerreviewmaketitle

\section{Introduction}
\label{sec:Introduction}


Decompilation is a critical technique that translates machine code (e.g., binary files) into human-readable languages~\cite{hex-rays, ghidra}. While decompiled code cannot be recompiled and executed, it significantly aids reverse engineers in comprehending, analyzing, and debugging programs when the source code is inaccessible. This makes decompilation indispensable in both software engineering~\cite{votipka2020observational} and cybersecurity~\cite{liu2022pg, ye2023cp}. Despite its importance, decompilation faces significant challenges due to factors such as the loss or absence of symbolic information, complex control flows, and other related issues, often leading to discrepancies between the decompiled code and the original source. These discrepancies, known as \textit{fidelity issues} (or \textit{distortion issues}), can severely affect both the readability and the semantic integrity of the decompiled code~\cite{dramko2024taxonomy}. As a result, decompiled code may suffer from issues such as meaningless variable names, type errors, redundant variables, incorrect return behavior, and the inclusion of compiler-specific functions. These problems not only hinder research but also pose a risk of misinterpreting the original code~\cite{burk2022decomperson, liu2020far}.

To this end, a few attempts have been made, aiming at improving the accuracy of source code inference from decompiled binaries, with a particular focus on predicting variable names and other code elements. For instance, Banerjee et al.~\cite{banerjee2021variable} introduced a method that employs masked language models, byte-pair encoding, and neural architectures to infer variable names in decompiled code. Similarly, HexT5~\cite{xiong2023hext5} leveraged a unified pre-training model with pseudo-code objectives, including code summarization and variable name recovery, while VARBERT~\cite{pal2024len} applied Transformer-based architectures to enhance variable name prediction in decompiled outputs. However, these machine learning-based approaches still have lots of limitations, particularly in that they can only target specific datasets and have limited generative capabilities. For example, these models often struggle to generate new code constructs or accurately infer variable names and types when encountering unfamiliar patterns or contexts. This limitation is particularly pronounced when dealing with code that exhibits significant deviations from the patterns seen during training, reducing their effectiveness in real-world scenarios where code diversity is usually higher than expected. 

In recent years, large language models (LLMs) have demonstrated outstanding performance in decompilation optimization. For instance, \degpt~\cite{hu2024degpt} was the first to use LLMs for optimizing the output of decompiled code. It introduced a three-role mechanism to enhance decompilation results, achieving significant improvements in restoring variable names and reconstructing code structures, thereby greatly enhancing the readability of decompiled code. However, this approach primarily addresses isolated aspects of decompilation fidelity, such as variable name recovery and code restructuring, leaving substantial gaps in fully detecting and correcting discrepancies between the source code and the decompiled outputs.

Overall, current research has yet to develop extensive and automated methods for detecting and addressing distortion issues. Generally, three primary challenges persist in mitigating this issue:

\textbf{C1: Impractical Description of Discrepancies.}
Existing research has primarily focused on identifying discrepancies in specific code elements, such as variables and types. However, these studies address only a small subset of the differences encountered during decompilation. While previous work \cite{dramko2024taxonomy} has provided a comprehensive taxonomy of these discrepancies, some types, such as missing variables and missing code, require reference to the source code for detection. This becomes impractical in closed-source environments where the source code is not directly available. Therefore, further investigation is necessary for closed-source scenarios.

\textbf{C2: Handling Long Decompiled Code in LLMs.} 
Decompiled functions often span hundreds of lines, posing a challenge for large language models (LLMs) due to token limitations and constrained attention mechanisms. Processing such long sequences in a single pass leads to performance degradation, increased computational costs, and loss of context for long-range variable dependencies. Efficiently managing long code sequences is critical for applying LLMs to large-scale decompilation tasks~\cite{wang2024hits,lu2024longheads}.


\textbf{C3: Inconsistent and Inaccurate Outputs of LLMs.}
One of the key challenges in optimizing decompilation using LLMs is ensuring accuracy and precision. LLMs often encounter issues such as semantic drift and hallucination during code interpretation, leading to inconsistent and unreliable results. Retrieval-Augmented Generation (RAG) can mitigate these problems by grounding LLM predictions with relevant external knowledge, but this requires a comprehensive database of distortion patterns and precise querying to ensure contextually accurate results.



In response to the challenges mentioned, we propose \tool, a novel framework designed to detect and correct distortions in decompiled code, especially in closed-source environments. This tool aims at optimizing decompiled output, providing reverse engineers with a more reliable foundation for analysis. Specifically, to address \textbf{C1}, we developed an extensive prompt template that systematically categorizes distortion types, aiding in the detection of decompilation issues in closed-source contexts(see \ref{sec:Distortion Types}). To tackle \textbf{C2}, we employ a chunking strategy to split long decompiled code into smaller, manageable blocks, reducing performance degradation in LLMs. To preserve context across chunks, we implement a Variable Dependency Algorithm(see \ref{sec:Variable Dependency Algorithm}) that extracts relationships between variables, ensuring accurate detection of distortions involving long-range dependencies.To solve \textbf{C3}, we leverage Retrieval-Augmented Generation (RAG) by constructing a decompilation distortion database containing annotated distortion patterns. We also apply a dynamic semantic intensity algorithm(see \ref{sec:Dynamic Semantic Intensity Retrieval Algorithm}) to identify potentially distorted code lines, enabling precise queries to the database. This enhances \tool’s ability to mitigate semantic drift and hallucination, improving the accuracy of distortion detection.

In this paper, we evaluate the performance of \tool across four key dimensions: distortion detection, distortion correction, algorithmic effectiveness (through ablation studies), and overall efficiency. First, in the distortion detection task, \tool demonstrates remarkable robustness, achieving an accuracy of 89\% and a precision of 83\%. These results highlight its strong capability in accurately identifying distortions. Second, for distortion correction, we introduce two metrics: Fix Rate and Corrected Fix Rate. The experimental results reveal that \tool achieves a Fix Rate of 94\% for detected distortions and a Corrected Fix Rate of 64\%, significantly outperforming baseline methods.
Third, we assess the effectiveness of the dynamic semantic intensity algorithm, which achieves an optimal balance between token usage and runtime performance. In addition, our variable dependency algorithm substantially reduces false negatives related to redundant code. The ablation study shows that the dynamic semantic intensity retrieval algorithm extracts more meaningful code lines, improving \tool's distortion detection performance, while the variable dependency algorithm achieves robust performance across different input code lengths. Finally, \tool exhibits exceptional efficiency, striking an excellent balance between token consumption and execution time, thus underscoring its practicality for real-world applications.

Our experiments on 620 function pairs from a binary similarity detection benchmark dataset~\cite{marcelli2022machine} validate \tool’s efficacy in addressing decompilation fidelity issues. By integrating LangChain for Retrieval-Augmented Generation (RAG) and leveraging a distortion database built from 150 examples in Dramko’s taxonomy~\cite{dramko2024taxonomy}, \tool pioneers the detection and correction of decompilation distortions, advancing reverse engineering in closed-source environments.

\begin{itemize}
    \item We created the first large-scale decompilation distortion dataset, with 620 function pairs and over 40,000 lines of code, enabling robust training and evaluation for distortion detection in closed-source settings.
    \item We propose \tool, a novel framework that uses RAG to detect and correct distortions, supported by a Decompilation Distortion Database of annotated distortion types and a Dynamic Semantic Intensity Algorithm to select semantically significant lines for efficient queries.
    \item We introduce the Variable Dependency Algorithm to preserve variable relationships across chunked decompiled code, addressing LLM context limitations and enhancing distortion detection accuracy.
    \item We evaluate \tool using Accuracy and Precision, and propose two new metrics, Fix Rate and Corrected Fix Rate, to measure distortion correction effectiveness, setting new standards for decompilation evaluation.
\end{itemize}

\section{Background \& Motivation}
\label{sec:Background}

Before presenting the technical details of \tool, we first review the foundations of decompilation (\ref{sec:decompilation}), discuss the RAG framework (\ref{sec:RAG}), and elaborate on our research motivations (\ref{sec:motivation}).

\subsection{Decompilation}
\label{sec:decompilation}

A decompiler, also known as a reverse compiler, is a tool that aims to reverse the compilation process. Given an executable program compiled from a high-level language, the decompiler seeks to generate a high-level language representation that approximates the functionality of the original program~\cite{cifuentes1995decompilation}. As software development and deployment have proliferated, decompilation has become increasingly critical in areas such as vulnerability discovery~\cite{mantovani2022convergence,chukkol2024vulcatch,reiter2022automatically,wang2023binvuldet}, malware analysis~\cite{mirzaei2021scrutinizer,almomani2022android,mauthe2021large}, and the comprehension of closed-source software~\cite{wang2023decompilation,pizzolotto2022bincc}. By decompiling, analysts can gain insights into the logic and behavior of a program, even in the absence of its original source code, enabling tasks such as debugging, vulnerability patching, and malware analysis. However, the decompilation process faces significant challenges due to the loss of high-level information during compilation, including variable names, comments, and structural elements of the source code. Despite these obstacles, decompilation remains an indispensable tool in security research and program analysis due to its ability to uncover insights from executable programs.

\subsection{Retrieval-Augmented Generation (RAG)}
\label{sec:RAG}

RAG is a framework that enhances generative models by integrating information retrieval. Before generating a response, the RAG model retrieves relevant documents or passages related to the input, which are then provided alongside the input to the generative model~\cite{zhao2024retrieval}. This approach enables the model to leverage external knowledge, improving the quality of the generated output. In contrast, traditional generative models, such as GPT, rely solely on the information encoded in the model’s parameters for text generation. This limits the model’s ability to generate accurate or up-to-date information, particularly when dealing with domain-specific queries or rapidly evolving topics~\cite{fan2024survey}. RAG addresses this limitation by incorporating a retrieval mechanism that fetches relevant documents or knowledge from external corpora during the generation process. One of the primary advantages of RAG is its ability to dynamically incorporate external knowledge without the need to retrain the underlying model. This is especially beneficial when the model needs to respond to domain-specific queries~\cite{wang2024domainrag} or handle real-time information~\cite{su2024dragin}. However, RAG also faces challenges, such as effectively selecting the most relevant documents from the corpus and balancing the retrieved information with the model’s inherent generative capabilities~\cite{zeng2024good}.

\subsection{Motivation}
\label{sec:motivation}

In this section, we detail our research motivation, aiming to tackle the raised challenges.

\subsubsection{Distortion Issues in a More Realistic Context}
\label{sec:Distortion Issues in a More Realistic Context}

Decompilation is a critical tool for reverse engineers analyzing software without access to source code. However, decompiled outputs often diverge significantly from the original source, introducing structural and semantic distortions that hinder analysis. Prior work, such as Dramko et al.'s taxonomy~\cite{dramko2024taxonomy}, has advanced understanding of fidelity issues but primarily focuses on open-source contexts where source code is available for validation. In closed-source scenarios, challenges like missing variables or entire code segments(see Section~\ref{sec:Case study I}) are often intractable, even with expert manual effort.

Given the inherent complexity of compiler-induced decompilation issues, which are often intractable without access to the original source code, our work shifts focus toward an application-driven perspective on fidelity issues. Rather than attempting to resolve all discrepancies, we prioritize those that are practically addressable in closed-source scenarios, ensuring actionable and relevant contributions to real-world decompilation.

\begin{figure*}[htbp]
    \centering
    \includegraphics[width=\textwidth]{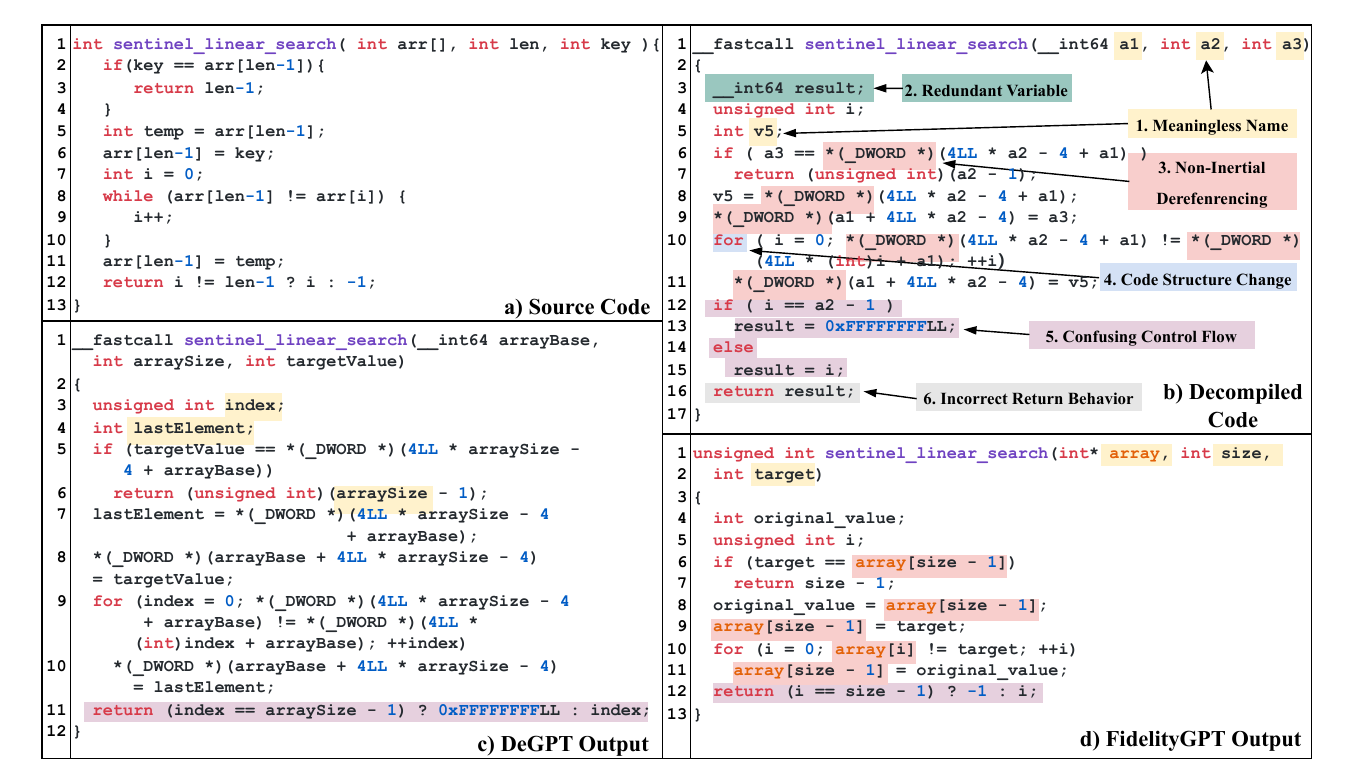}
    \caption{Case Study of Decompilation Fidelity Issues and Comparison of \tool with DeGPT}
    \label{fig:motivation2}
\end{figure*}

Figure~\ref{fig:motivation2} provides a case study to illustrate decompilation fidelity issues. It is important to note that \textbf{these discrepancies are strictly based on a comparison between the decompiled code and the corresponding source code}. Panel (a) presents the source code, while panel (b) displays the corresponding decompiled code. The following differences are notable:

\begin{enumerate}[left=0em]
    \item \textbf{Meaningless Parameter and Variable Names:} The source code features descriptive names that facilitate understanding, whereas the decompiled code often uses ambiguous names, complicating comprehension.
    
    \item \textbf{Redundant Variables:} The source code uses two variables, \texttt{temp} and \texttt{i}, whereas the decompiled code includes an additional, redundant variable, \texttt{result} without adding any functional significance.
    
    \item \textbf{Non-Inertial Dereferencing:} In the source code, array types are accessed directly. However, the decompiled code accesses arrays via pointers, with the representation of array members being obscure and difficult to interpret.
    
    \item \textbf{Code Structure Changes:} The source code employs a \texttt{while} loop, whereas the decompiled code uses a \texttt{for} loop, indicating structural differences.
    
    \item \textbf{Confusing Control Flow:} The source code uses a ternary operator for value retrieval, while the decompiled code replaces it with \texttt{if} statements, leading to a more complex and less intuitive control flow.
    
    \item \textbf{Incorrect Return Behavior:} The source code uses a ternary operator to determine return values, while the decompiled code returns values directly, creating inconsistencies.
\end{enumerate}

These distortions impair readability and correctness, hindering security analysts’ ability to identify vulnerabilities. For example, ambiguous names reduce comprehension, while pointer-based access may mask errors, risking missed threats. Prior approaches like DeGPT improve variable renaming but fail to address redundant variables and pointer issues, as shown in panel (c) of Figure~\ref{fig:motivation2}. In contrast, panel (d) shows \tool’s output, leveraging a distortion database, variable dependency analysis, and semantic intensity scoring to enhance clarity and accuracy, significantly improving binary analysis for security professionals.

\subsubsection{Overcoming Long-Range Variable Dependencies}

\begin{figure*}[htbp]
    \centering
    \includegraphics[width=\textwidth]{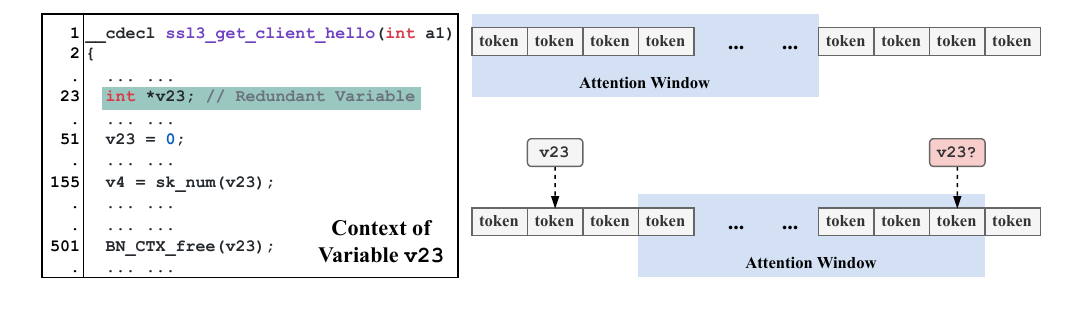}
    \caption{Long-Range Variable Dependencies in Decompiled Code}
    \label{fig:motivation1}
\end{figure*}

Due to the inherent limitations of Large Language Models (LLMs), such as restricted input window sizes and constrained attention mechanisms, processing large volumes of decompiled code in a single pass significantly impairs performance in decompilation distortion detection tasks. The limited input window restricts the amount of code that can be analyzed simultaneously, while the attention mechanism struggles to maintain contextual connections across lengthy code sequences, leading to reduced distortion detection accuracy. As shown in Figure~\ref{fig:motivation1}, the definition of \texttt{v23} at line 23 and its usage at line 501, separated by over 400 lines, may exceed the attention window, degrading detection performance. To address these limitations, we propose partitioning the decompiled code into smaller, manageable chunks. This chunking strategy reduces input complexity, allowing the LLM to operate within its input window and alleviating the burden on the attention mechanism, thereby enhancing analytical performance.

However, chunking risks splitting long-range variable dependencies across chunks, which could prevent the LLM from detecting distortions involving these variables. For instance, in Figure~\ref{fig:motivation1}, the definition of \texttt{v23} at line 23 and its usage at line 501 may be split into different chunks, depriving the LLM of the context needed to identify variable redundancy. To overcome this, we propose extracting dependency relationships between variables in the decompiled code. By explicitly capturing these dependencies, we enable systematic analysis of variable redundancy across chunks, ensuring the LLM can accurately assess redundancy. This combined approach of code chunking and variable dependency extraction improves the precision and reliability of decompilation distortion detection, effectively mitigating the constraints of input window size and long-range dependencies.

\subsubsection{Mitigating Semantic Drift and Hallucination in Distortion Detection via RAG}

Large Language Models (LLMs) often produce inaccurate outputs in decompilation tasks due to \textit{semantic drift}, where they misinterpret the context of code structures, and \textit{hallucination}, where they generate erroneous elements. To address this, we employ Retrieval-Augmented Generation (RAG) to ground LLM predictions with contextually relevant knowledge, enhancing the accuracy of distortion detection. However, effective RAG deployment requires addressing two challenges: constructing a diverse distortion database and formulating precise input queries.

First, we curate a decompilation distortion database of distortion instances annotated with specific distortion types (e.g., redundant variables, control flow obfuscation), ensuring coverage of diverse decompilation scenarios. This diversity enables RAG to retrieve semantically aligned examples, improving the model’s capacity to discern subtle distortion patterns. Second, formulating precise input queries is essential for targeted retrieval. We feed RAG with carefully selected code lines that exhibit potential distortion patterns. This approach ensures the retrieved knowledge directly informs the LLM’s analysis, refining its ability to distinguish genuine distortions from hallucinated artifacts.

By addressing these challenges, \tool utilizes RAG to mitigate LLM hallucinations, ensuring reliable and accurate distortion detection in decompiled code.





\section{Methodology}
\label{sec:Methodology}

Our primary objective is to extensively detect and address discrepancies between decompiled code and source code within closed-source environments, with a particular emphasis on fidelity issues. We aim to fix decompiled code, making it closer to the original source code format, thereby aiding reverse engineers in analyzing binary files. In this section, we introduce the design of \tool, a framework built upon a large language model that leverages Retrieval-Augmented Generation (RAG) techniques to enhance the interpretability of decompiler outputs for security analysts.

We begin by providing an overview of the workflow, followed by discussing the challenges associated with accurately describing fidelity issues and the technical complexities inherent in utilizing LLMs and RAG. Finally, we detail the design of the prompt templates employed in the tasks of detection and correction.

\subsection{Overview}
\label{sec:overview}

\begin{figure*}[htbp]
    \centering
    \includegraphics[width=\textwidth]{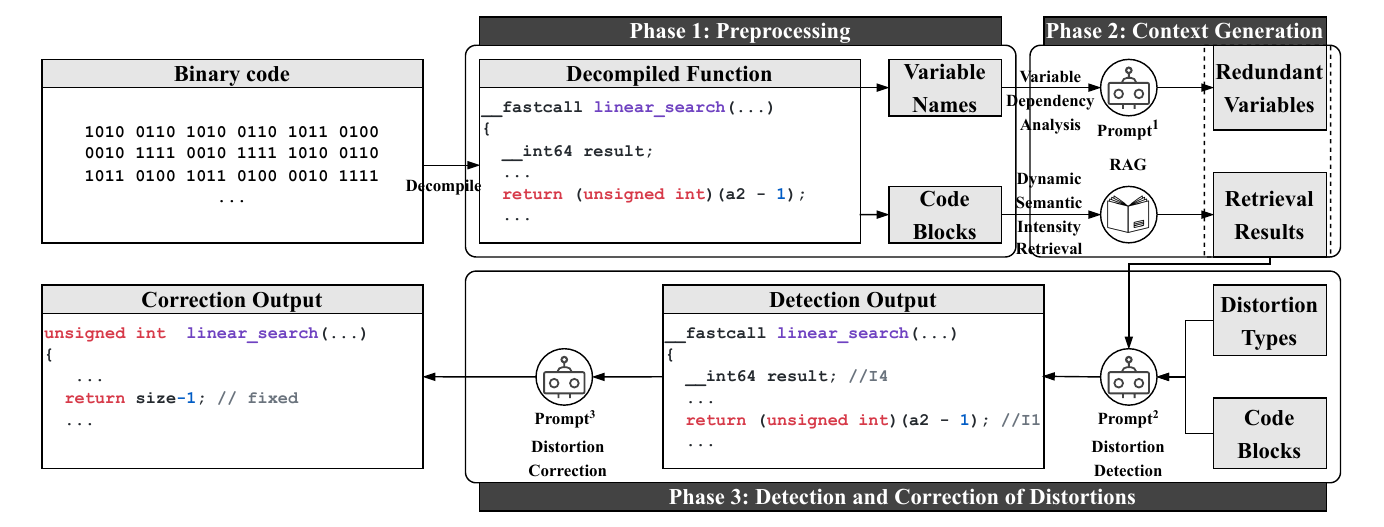}
    \caption{Workflow of \tool}
    \label{fig:workflow}
\end{figure*}
As illustrated in Fig.~\ref{fig:workflow}, \tool operates through three main phases: \textbf{Preprocessing}, \textbf{Context Generation}, and \textbf{Detection and Correction of Distortions}, each designed to address the discrepancies between decompiled and source code. 

\textbf{Phase 1: Preprocessing.} In this phase, the binary code is first decompiled into its corresponding functions. If any \textit{decompiled function} exceeds a predefined line threshold (see \ref{sec:Effectiveness of Algorithm}), it is segmented into smaller manageable blocks. A divide-and-conquer approach is applied, dividing the decompiled function into \textit{variable names} and \textit{code blocks}. This segmentation ensures that larger functions can be processed efficiently in later stages.

\textbf{Phase 2: Context Generation.} Two novel methods generate context for distortion detection. First, the \textbf{Variable Dependency Algorithm}~\ref{sec:Variable Dependency Algorithm} extracts dependency relationships for variables identified in preprocessing. These are fed into \textit{Prompt Template 1}~\cite{fidelitygpt} to flag redundant variables via LLM analysis. Second, the \textbf{Dynamic Semantic Intensity Retrieval Algorithm}~\ref{sec:Dynamic Semantic Intensity Retrieval Algorithm} selects semantically significant lines and queries a Decompilation Distortion Database using Retrieval-Augmented Generation (RAG) to retrieve relevant code. Redundant variables and retrieved code form \textit{Prompt Template 2} for distortion detection in the next phase.

\textbf{Phase 3: Detection and Correction of Distortions.} In this final phase, predefined \textbf{distortion types} are combined with the \textit{code blocks} and the \textit{context} generated in the previous phase to form \textit{Prompt Template 2}(see Fig.~\ref{fig:prompt} panel(a)), which is used for distortion detection via the LLM. The output of this step is a decompiled function annotated with specific \textit{distortion type identifiers} (e.g., `// I4`(see \ref{sec:Distortion Types})). The identified distortions are then passed into \textit{Prompt Template 3}(see Fig.~\ref{fig:prompt} panel(b)), which generates the corrected code annotated with '\textit{// fixed}' markers, providing clear visibility of the modifications made to resolve the distortions.

\subsection{Variable Dependency Algorithm}
\label{sec:Variable Dependency Algorithm}

The attention mechanism of LLMs struggles to process long decompiled code in a single pass, often missing critical distortions such as redundant variables (see Section~\ref{sec:Effectiveness of Algorithm}). To address this, we adopt a divide-and-conquer approach, segmenting decompiled code exceeding a predefined threshold into smaller, manageable chunks. This chunking strategy preserves analysis consistency but introduces a challenge: variable definitions and their usages may be split across chunks, leading to false positives in redundant variable detection due to incomplete contextual information.

Our analysis indicates that the decompilation process frequently introduces redundant variables due to register reuse patterns. To systematically detect these variables, \textit{Prompt Template 1} defines the following formal criteria for redundancy identification:

\begin{enumerate}
\item \textbf{Temporary Variables}: Variables used solely for short-term data storage. These are typically introduced during decompilation to hold transient values, such as register contents, and are not referenced beyond their immediate context.
\item \textbf{Intermediate Variables}: Variables employed in intermediate computational steps. These variables store results of temporary calculations, often generated by decompilation processes, and are only relevant within specific operations.
\item \textbf{Duplicate Variables}: Variables that replicate data already represented by other variables or constants. Such variables arise from redundant assignments or register reuse, unnecessarily duplicating information.
\item \textbf{Low-Usage Variables}: Variables referenced infrequently within the code. These variables, often used once or twice, do not contribute significantly to the program's logic and may indicate redundancy.
\item \textbf{Non-Significant Variables}: Variables lacking independent semantic importance. These variables do not convey unique information and are typically artifacts of decompilation without meaningful roles.
\item \textbf{Mergeable Variables}: Variables whose data can be logically combined with other statements. These variables store information that can be integrated into existing statements, reducing code complexity.
\end{enumerate}

By integrating the \textbf{Variable Dependency Algorithm} with \textit{Prompt Template 1}, we enable the LLM to identify redundant variables based on their dependency relationships and these criteria.

The \textbf{Variable Dependency Algorithm} (Algorithm~\ref{alg:variable_dependency}) generates a mapping $\mathcal{M}$ of variables to their dependent statements in decompiled code, enabling detection of redundant variables across chunked code. It operates in two steps: constructing a Program Dependence Graph (PDG) and tracing variable dependencies.

First, the \texttt{VariableDependencyAnalysis} function builds the PDG by parsing the decompiled code $C$ into variables (\texttt{var}) and statements (\texttt{lines}). It constructs a Control Flow Graph (CFG), computes control dependencies (CDG) and data dependencies (DDG), and combines them into the PDG to capture variable interactions.

Second, the \textsc{TraceVariable} procedure traces dependencies for each variable \texttt{var} by traversing the PDG. It collects statements from \texttt{lines} that depend on \texttt{var} into a list $\mathcal{D}$, using a set of visited nodes $\mathcal{V}$ to avoid redundant traversals, and recursively traces related variables. The mappings $\mathcal{M}$ store these dependencies.

These mappings are integrated into \textit{Prompt Template 1}~\cite{fidelitygpt}, which defines redundancy criteria, and fed to the LLM to detect redundant variables. This approach addresses chunking and LLM attention limitations, improving the accuracy and efficiency of redundant variable detection in decompiled code.

\begin{algorithm}[htbp]
\footnotesize 
\SetAlgoNlRelativeSize{0}
\caption{Variable Dependency Algorithm}\label{alg:variable_dependency}
\KwData{Decompiled code $C$}
\KwResult{Variable dependency mappings $\mathcal{M}$: a mapping of variables to their dependent statements}
\Comment{$\mathcal{G}$: Program Dependence Graph (PDG), $var$: variable name, $lines$: list of code statements, $\mathcal{V}$: set of visited nodes, $\mathcal{D}$: list of dependent statements}

\SetKwFunction{FVarDepAnalysis}{VariableDependencyAnalysis}
\SetKwProg{Fn}{Function}{:}{}
\SetKwProg{Proc}{Procedure}{:}{}

$\mathcal{M} \gets \FVarDepAnalysis{C}$ 
\KwRet{$\mathcal{M}$}

\Fn{\FVarDepAnalysis{$C$}}{
    \texttt{var} $\gets$ parse $C$ to identify all variables \\
    \texttt{lines} $\gets$ split $C$ into statements \\
    \texttt{cfg} $\gets$ build CFG from statements \\
    \texttt{cdg} $\gets$ compute control dependencies \\
    \texttt{ddg} $\gets$ compute data dependencies \\
    \texttt{pdg} $\gets \texttt{cdg} \cup \texttt{ddg}$ \\
    $\mathcal{M} \gets \{\}$ 
    \For{$variable \in \texttt{var}$}{
        $\mathcal{V} \gets \emptyset$, $\mathcal{D} \gets [\,]$ 
        \Proc{\textsc{TraceVariable}($x$)}{
            \For{$predecessor \in \texttt{pdg}.\text{pred}(x)$}{
                \If{$predecessor \notin \mathcal{V}$}{
                    $\mathcal{V} \gets \mathcal{V} \cup \{predecessor\}$ 
                    \If{$predecessor \in \mathrm{dom}(\texttt{lines})$}{
                        $\mathcal{D} \gets \mathcal{D} \cup \{\texttt{lines}[predecessor]\}$ 
                        \For{$dep\_variable \in \text{var}(\texttt{lines}[predecessor])$}{
                            \textsc{TraceVariable}($dep\_variable$) 
                        }
                    }
                }
            }
        }
        \textsc{TraceVariable}($var$)
        $\mathcal{M}[var] \gets \mathcal{D}$ 
    }
    \KwRet{$\mathcal{M}$}
}
\end{algorithm}

\subsection{Dynamic Semantic Intensity Retrieval Algorithm}
\label{sec:Dynamic Semantic Intensity Retrieval Algorithm}

Directly feeding entire decompiled functions or code blocks into a Retrieval-Augmented Generation (RAG) system for similarity-based retrieval from the Decompilation Distortion Database is often inefficient. The high similarity among code segments frequently results in redundant retrievals, wasting computational resources and hindering the reasoning capabilities of LLMs.

The \textbf{Dynamic Semantic Intensity Retrieval Algorithm} (Algorithm~\ref{alg:semantic_intensity}) selects the top-$k$ semantically significant lines from decompiled code to enhance Retrieval-Augmented Generation (RAG) for distortion detection. It scores lines based on syntactic constructs and retrieves diverse, high-intensity lines to query the Decompilation Distortion Database $\mathcal{D}$, mitigating LLM semantic drift and hallucination.

The \texttt{GenerateSemanticIntensityLines} function assigns intensity scores to each line in $lines$ using $feature\_weights$ derived from frequency analysis of constructs (e.g., assignments, loops, function calls) in $\mathcal{D}$. Scores are stored in $intensities$ as line-intensity pairs and sorted descendingly. The number of selected lines, $k$, is set to $total\_lines$ if below $min_lines$, or $\min(base\_lines + \lfloor \frac{total\_lines - threshold}{step} \rfloor, max\_lines)$ otherwise, prioritizing diverse constructs.

These $selected_lines$ query $\mathcal{D}$ to retrieve similar code, guiding the LLM to detect distortions accurately while reducing computational overhead.

\begin{algorithm}[htbp]
\footnotesize 
\SetAlgoNlRelativeSize{0}
\caption{Dynamic Semantic Intensity Retrieval Algorithm}\label{alg:semantic_intensity}
\KwData{Decompiled code $C$, Decompilation Distortion Database $\mathcal{D}$}
\KwResult{Top-$k$ semantically significant lines}
\Comment{$lines$: List of code lines from $C$ to analyze, $min\_lines$: Minimum number of lines to select, $base\_lines$: Base number of lines to select, $threshold$: Threshold for total lines to adjust selection, $step$: Step size for dynamic line selection, $max\_lines$: Maximum number of lines to select, $\mathcal{D}$: Database with frequency data for syntax constructs.}

\SetKwFunction{FProcessLines}{GenerateSemanticIntensityLines}
\SetKwProg{Fn}{Function}{:}{}

\Fn{\FProcessLines{$lines$, $min\_lines$, $base\_lines$, $threshold$, $step$, $max\_lines$, $\mathcal{D}$}}{
    $feature\_weights \gets \text{weights from frequency analysis of syntax constructs in } \mathcal{D}$ \tcp*{Compute weights for syntactic constructs}
    $intensities \gets \text{empty list}$
    
    \For{each line in $lines$}{
        \If{line contains constructs \{assignment, addition, variable definition, return, loop, conditional, function call, type\}}{
            $intensity \gets \text{sum of } feature\_weights \text{ for detected constructs}$ 
            $intensities.append((line, intensity))$ 
        }
    }
    
    \textbf{Sort} $intensities$ by intensity in descending order 
    
    \If{$total\_lines \leq min\_lines$}{
        $k \gets total\_lines$ 
    }
    \Else{
        $k \gets \min(base\_lines + \lfloor \frac{total\_lines - threshold}{step} \rfloor, max\_lines)$ 
    }
    $selected\_lines \gets \text{top } k \text{ lines, prioritizing diverse construct types}$ \tcp*{Select top-$k$ lines with varied constructs}
    \KwRet{$selected\_lines$}
}

\end{algorithm}

\subsection{Distortion Types}
\label{sec:Distortion Types}
DeGPT~\cite{hu2024degpt} is currently a leading framework for optimizing decompiler outputs. It excels at addressing variable renaming, simplifying code structures, and generating meaningful comments, significantly enhancing the efficiency of security analysts. However, the differences between source code and decompiled code go beyond variables and code structure. To comprehensively address these discrepancies, we adopt a predefined fidelity taxonomy~\cite{dramko2024taxonomy}. As previously mentioned(see \ref{sec:motivation}), this taxonomy, originally proposed in the context of open-source environments, contrasts with the typical scenario in which reverse engineers perform decompilation without access to the source code. Therefore, we redefine decompilation fidelity issues in closed-source environments. In this context, we carefully selected and integrated specific fidelity types to enable the detection of source code versus decompiled code differences directly through analysis of the decompiled code. We have designed six classifications of fidelity issues, labeled I1 to I6, as detailed below.

\begin{enumerate}[left=0em,label=\textbf{I\arabic*:}]
    \item \textbf{Non-inertial dereferencing.} Accessing structure members through pointers and arrays, or accessing array members using pointers.
    \item \textbf{Character and string literal representation issues.} String literals being replaced with references or represented as integers.
    \item \textbf{Control flow obfuscation.} Transformations involving while and for loops and destructuring ternary operators.
    \item \textbf{Redundant code.} Includes redundant variables, meaningless parameter assignments, variable assignments within functions without return values, and redundant variables introduced by non-inertial dereferencing.
    \item \textbf{Unexpected returns.} Function structures or return values that deviate from expected outcomes.
    \item \textbf{Use of non-type symbols.} Employing symbols and macros that do not match types in decompiled code, using symbols and user macros or function calls that do not match types in semantically equivalent decompiled code, and including compiler-specific functions.
\end{enumerate}

In summary, this detailed characterization offers an extensive understanding of decompilation fidelity issues within closed-source environments. It establishes a solid foundation for the subsequent tasks of detection and correction.

\subsection{Prompting for Distortion Detection and Correction}
\label{sec:Detection and Correction}
The final stage comprises two key tasks: \textbf{distortion detection} and \textbf{correction}. The detection task prompt template primarily defines the various distortion types. In contrast, the correction task prompt template outlines methods for correcting these distortions, ultimately leading to an optimized version of the decompiled code.

\begin{figure*}[htbp]
    \centering
    \includegraphics[width=\textwidth]{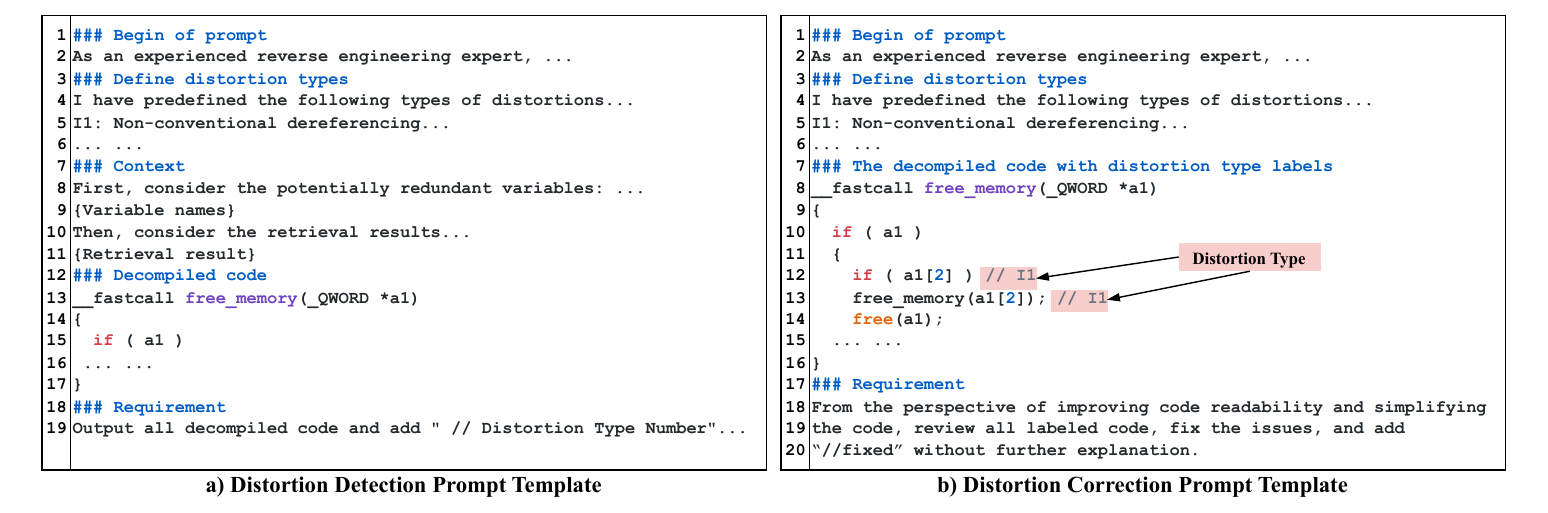}
    \caption{Prompt Templates}
    \label{fig:prompt}
\end{figure*}

In constructing the \textbf{distortion detection prompt template}, we carefully considered several aspects to ensure it effectively supports distortion detection in decompiled code. The prompt, as shown in panel (a) of Fig.~\ref{fig:prompt}, consists of the following key components:

\begin{itemize}[left=0em]
    \item \textbf{Role and Task Definition.} At the beginning of the template, we explicitly define the user as an "experienced reverse engineering expert." This sets the context for the model, preparing it to assume the role of a highly skilled professional. 

    \item \textbf{Predefined Distortion Types.} To systematically identify distortions, we introduce six predefined distortion types labeled as \textbf{I1} through \textbf{I6}. These types serve as a reference framework for consistent classification during analysis, mitigating the risk of subjective bias. 

    \item \textbf{Context.} To enhance the model's understanding and analysis of decompiled code, we integrated redundant variable names with the results retrieved from RAG. These results serve as the context for the language model, facilitating more accurate identification of distortions during decompilation.

    \item \textbf{Decompiled Code and Required Output.} The core section of the template involves the decompiled code itself. We specify the format for the model’s output, requiring it to annotate the detected distortion types directly within the code, using the labels defined earlier (\textbf{I1--I6}). This ensures that the output is structured for ease of evaluation and analysis.
\end{itemize}

Similarly, as shown in panel (b) of Fig.~\ref{fig:prompt}, the design of the \textbf{distortion correction prompt template} follows the same principles as the detection template. The major difference is in the output format. While the detection template focuses on identifying and labeling distortion types, the correction template requires the model to fix the distortions and mark the corrected lines with a \textbf{"//fixed"} annotation. This approach allows for a direct comparison between the detected distortions and their corresponding repairs, enabling a clear assessment of the effectiveness of the model’s correction process.

\section{Evaluation}
\label{sec:Evaluation}

In this section, we evaluate our approach in order to answer the following research questions.

\begin{itemize}[left=0em]
  \item \textbf{RQ1:} How effective is \tool in distortion detection?
  \item \textbf{RQ2:} How effective is \tool in distortion correction?
  \item \textbf{RQ3:} How does the impact of each component of \tool on its overall effectiveness?
  \item \textbf{RQ4:} How is the efficiency of \tool ?
  \item \textbf{RQ5:} How robust and generalizable is \tool across settings?

\end{itemize}

\subsection{Implementation}


\tool utilizes the \textit{GPT-4o} API with a temperature setting of \textit{0.5}. The RAG component is implemented using \textit{LangChain}, with the embedding model set to \textit{text-embedding-ada-002}. The retrieval method is similarity-based, with \(k = 1\).

To ensure effective RAG operation, we identify three necessary design conditions:

\textbf{(1) Informative query selection.}  
We apply the \textit{Dynamic Semantic Intensity Retrieval Algorithm} to extract the most semantically intense lines from the input function—i.e., those most likely to exhibit distortion. These selected lines are used as targeted queries to the RAG database, improving retrieval relevance and downstream detection performance by reducing noise from less informative lines.

\textbf{(2) Line-level similarity retrieval.}  
Each selected line is embedded using \texttt{text-embedding-ada-002}, and the most semantically similar distorted line is retrieved from the database. This fine-grained, line-level matching ensures structural and contextual alignment with real-world distortion cases. Importantly, the retrieved lines are annotated with distortion types, providing actionable semantic cues for the LLM to infer distortion presence.

\textbf{(3) Decompiled distortion database.}  
We constructed a domain-specific distortion database by filtering and annotating distorted code lines from the public dataset introduced in~\cite{dramko2024taxonomy}. All entries were deduplicated and labeled at the line level with fine-grained distortion types (\textit{I1}–\textit{I6}). To ensure diversity, the database includes 150 lines derived from \texttt{IDA Pro} and 91 additional lines from \texttt{Ghidra} decompilation outputs. This manually curated resource provides the retrieval foundation for distortion-aware reasoning and supports cross-tool generalizability.

Together, these components form a targeted and interpretable RAG pipeline that equips the LLM with reliable, context-rich priors grounded in realistic distortion cases.

\subsection{Dataset \& Setup}\label{evaluation:dataset_&_setup}

\subsubsection{Dataset}\label{evaluation:dataset_&_setup:dataset}

To evaluate our approach for detecting distortions in decompiled code, we repurposed a subset of functions from established binary similarity detection datasets~\cite{marcelli2022machine}, including \textit{Coreutils-ARM-32}, \textit{Curl-MIPS-32}, \textit{ImageMagick-ARM-32}, \textit{OpenSSL-X86-32}, \textit{Putty-X86-32}, and \textit{SQLite-X86-32}. These benchmarks were selected for their quality, architectural diversity, and the availability of source–binary mappings. To enhance diversity, we also included the \textit{CAlgorithm-X86-64} repository~\cite{TheAlgorithms_C}, which contains community-written algorithms with varying coding styles and complexity.

From these sources, we randomly selected 55, 110, 60, 150, 55, 90, and 100 pairs of decompiled functions and their corresponding source code, based on the number of decompiled functions extracted from each project. All functions were decompiled using \textit{IDA Pro 7.5} under the \texttt{-O0} optimization level to ensure maximal alignment with the source code. We then manually annotated the distortion types (\textit{I1} to \textit{I6}) for each function pair using a consistent labeling protocol, resulting in over 40,000 lines of annotated code.

To evaluate the generalization of \tool across compilers and optimization levels, we further constructed datasets using \texttt{-O1}, \texttt{-O2}, and \texttt{-O3} binaries (decompiled with \textit{IDA Pro}) and a \texttt{-O0} dataset decompiled using \textit{Ghidra}. All were annotated following the same protocol.

Please refer to Appendix~\ref{sec:dataset_analysis_appendix} for detailed statistics and representativeness analysis of the evaluation subset.

\subsubsection{Algorithm Configuration}
The threshold for partitioning decompiled code blocks is set at \textit{50} lines. The Dynamic Semantic Strength Retrieval Algorithm is designed to output between \textit{5} and \textit{10} lines of code, depending on the complexity and length of the input. The parameters used to control the output are as follows: \textit{min\_lines = 5}, \textit{base\_lines = 5}, \textit{max\_lines = 10}, \textit{threshold = 5}, and \textit{step = 9}.

\subsubsection{Metrics}

In the distortion detection task, we primarily use accuracy (Acc) as the evaluation metric. Additionally, since false positives (FP) may impact downstream distortion correction tasks(see \ref{sec:Case study III}), we also include precision (Pr) as an additional metric. The specific calculation methods are as follows:

\begin{equation}
\footnotesize
\text{Acc} = \frac{\text{TP} + \text{TN}}{\text{TP} + \text{TN} + \text{FP} + \text{FN}}, \quad
\text{Pr} = \frac{\text{TP}}{\text{TP} + \text{FP}}
\end{equation}

where TP (True Positive) refers to the number of correctly identified distortions.
TN (True Negative) refers to the number of correctly identified non-distortions.
FP (False Positive) refers to the number of non-distortions incorrectly identified as distortions.
FN (False Negative) refers to the number of distortions that were missed or incorrectly identified as non-distortions.

In the distortion correction task, we evaluate \tool from the perspectives of alignment with the fidelity definitions, assessing from the perspectives of correctness and readability. We utilize two manually computed metrics, \textit{ Fix Rate (FR)} and \textit{Corrected Fix Rate (CFR)}, similar to the evaluation of \degpt~\cite{hu2024degpt}.

\noindent\textbf{FixRate (FR)} measures the improvement in code semantics restoration and readability. It inspects all lines marked with "\textit{//fixed}" to assess the effectiveness of distortion correction. If the identified distortion is rectified by comparing it with the decompiled code, the code is considered fixed.

\begin{equation}
\footnotesize
\begin{aligned}
\text{FR} &= \frac{\text{Fixed lines of code}}{\text{all annotations fixed lines of code}}, \\
\text{CFR} &= \frac{\text{Corrected fixed lines of code}}{\text{all annotations fixed lines of code}}
\end{aligned}
\end{equation}

\noindent\textbf{Corrected Fix Rate (CFR)} differs from FR in that it focuses solely on correctness issues. Since variable renaming and type changes alone can improve readability without altering code semantics, they are not considered as correctness fixes. Therefore, CFR emphasizes semantic corrections. The specific evaluation criteria are detailed in the table\ref{tab:classification_fixes}.

\begin{table*}[ht]
\centering
\footnotesize
\caption{Classification of Fixes for Variable and Code Corrections.}
\begin{tabular}{|c|c|c|c|c|c|c|}
\hline
\textbf{\makecell{Fixed \\ Content}}  & \textbf{\makecell{Variable \\ Renaming}} & \textbf{\makecell{Variable Type \\ Modification}} & \textbf{\makecell{Variable Type \\ Restoration}} & \textbf{\makecell{Numerical and \\ Character}} & \textbf{\makecell{Code \\ Structure}} & \textbf{\makecell{Non-Type \\ Symbols}} \\ \hline
\textbf{FR} &  \makecell{\checkmark} & \makecell{\checkmark} & \makecell{\checkmark} & \makecell{\checkmark} & \makecell{\checkmark} & \makecell{\checkmark} \\ \hline
\textbf{CFR} &  & & \makecell{\checkmark} & \makecell{\checkmark} & \makecell{\checkmark} & \makecell{\checkmark} \\ \hline
\end{tabular}
\label{tab:classification_fixes}
\end{table*}

Assessing the correctness of automatic correction results is challenging. Consequently, we manually compute FR and CFR during the evaluation process to ensure reliability. Higher FR and CFR values indicate better correction performance.

\subsection{Baselines}

This section provides a brief explanation of the baseline methods used in our evaluation:
\begin{itemize}[left=0em]
\item \textbf{\promptz (zero-shot):} The distortion correction prompt template excludes the definition of distortion types and distortion labels. Since it does not include distortion definitions, this method serves as a comparison for directly using the LLM for distortion correction.

\item \textbf{\promptd (with definitions):} The distortion detection prompt template excludes contextual information. However, it includes the definition of distortion types, making it applicable for both distortion detection and correction phases.

\item \textbf{\prompte (with examples):} The distortion detection prompt template excludes contextual information but includes three additional examples. This method also contains distortion type definitions, and uses them for comparison in both distortion detection and correction phases.

\item \textbf{\degpt~\cite{hu2024degpt}:} This method represents the current state-of-the-art (SOTA) in decompilation optimization, leveraging the GPT-4o API for its implementation. It focuses on three key tasks: variable renaming, code structure simplification, and comment generation. As our work does not involve comment generation, our comparison centers on variable renaming and code structure simplification to evaluate the optimization capabilities of DeGPT. Similarly, since it does not include distortion definitions, this method serves as a baseline for distortion correction.

\item \textbf{\textit{LLM4Decompile}}~\cite{tan2024llm4decompile}: This method refines decompiled code from traditional decompilers like Ghidra to generate high-level source code, such as C, using a large language model fine-tuned on DeepSeek-Coder with approximately 40 billion tokens of assembly-to-C code pairs. As our work focuses on optimizing decompiled code output, we compare with LLM4Decompile-Ref.

\item \textbf{\textit{ReSym}}~\cite{xie2024resym}: This approach combines large language models with program analysis to recover variable names and field accesses from stripped binary files. ReSym fine-tunes two models, VarDecoder and FieldDecoder, for predicting variable names and field accesses, respectively, and uses a Prolog-based algorithm to reduce uncertainty in LLM outputs. While effective in symbolic restoration, ReSym does not focus on optimizing the structural or functional aspects of decompiled code.

\end{itemize}

\subsection{RQ1: How effective is \tool in distortion detection?}
\label{sec:Distortion detection experiment}
Since \degpt does not include distortion detection functionality, we use \promptd and \prompte as baselines. Table \ref{tab:detection} summarizes the distortion detection results for various approaches, including \promptd, \prompte, and \tool. The evaluation focuses on two key metrics: Accuracy and Precision. Results are provided for each individual dataset, as well as the overall average across all datasets.

As shown in Table \ref{tab:detection}, in terms of accuracy, \tool achieved the highest average accuracy of 0.89, outperforming both \promptd and \prompte across all datasets. \prompte and \promptd methods performed similarly, with an average accuracy of 0.88. Regarding precision, \tool again demonstrated superior performance with an average precision of 0.83, significantly surpassing the \promptd (0.73) and the \prompte (0.75). This suggests that \tool is more effective at minimizing false positives while maintaining high accuracy in detecting distortions. The \prompte method showed higher average precision than \promptd, this indicates that the inclusion of examples helps refine precision.

From the experimental results, it is evident that all methods achieved similar high accuracy due to the use of prompt templates based on systematically defined distortions. However, \tool demonstrated superior precision. The integration of RAG significantly reduced false positives, making \tool more reliable in accurately detecting distortions. The combination of efficient design and the precision of RAG highlights the distinct advantages of our tool.

\begin{table*}[htbp]
\centering
\caption{Performance Comparison of Different Approaches on Distortion Detection across Various Datasets.}
\scriptsize
\resizebox{\textwidth}{!}{  
\begin{tabular}{|l|c|c|c|c|c|c|c|c|c|c|c|c|c|c|c|c|}
\hline
\textbf{Approach}  & \multicolumn{2}{c|}{\textbf{ImageMagick}} & \multicolumn{2}{c|}{\textbf{curl}} & \multicolumn{2}{c|}{\textbf{putty}} & \multicolumn{2}{c|}{\textbf{CAlgorithm}} & \multicolumn{2}{c|}{\textbf{coreutils}} & \multicolumn{2}{c|}{\textbf{OpenSSL}} & \multicolumn{2}{c|}{\textbf{SQLite}} & \multicolumn{2}{c|}{\textbf{Average}} \\ \hline
  \textbf{Metrics} & \textbf{Acc} & \textbf{Pr} & \textbf{Acc} & \textbf{Pr} & \textbf{Acc} & \textbf{Pr} & \textbf{Acc} & \textbf{Pr} & \textbf{Acc} & \textbf{Pr} & \textbf{Acc} & \textbf{Pr} & \textbf{Acc} & \textbf{Pr} & \textbf{Acc} & \textbf{Pr} \\ \hline
\textbf{\promptd}  & 0.89 & 0.66 & 0.90 & 0.85 & 0.88 & 0.78 & 0.88 & 0.66 & 0.82 & 0.67 & 0.87 & 0.62 & 0.88 & 0.86 & 0.87 & 0.73 \\ \hline
\textbf{\prompte}  & 0.91 & 0.74 & 0.88 & 0.87 & 0.86 & 0.75 & 0.88 & 0.69 & 0.84 & 0.67 & 0.88 & 0.64 & 0.88 & 0.87 & 0.88 & 0.75 \\ \hline
\textbf{\tool}  & 0.91 & 0.83 & 0.90 & 0.88 & 0.89 & 0.85 & 0.90 & 0.71 & 0.85 & 0.79 & 0.89 & 0.81 & 0.88 & 0.92 & 0.89 & 0.83 \\ \hline
\end{tabular}
}
\label{tab:detection}
\end{table*}





\noindent\colorbox{gray!20}{
    \parbox{\dimexpr\linewidth-2\fboxsep}{
    \textbf{Answer to RQ1:} From the experimental results, it is evident that \tool achieved the highest accuracy and precision across all datasets. This demonstrates that \tool performs effectively in the distortion detection task. Furthermore, the RAG retrieval results as prompts yielded favorable outcomes, further enhancing the performance of \tool.
    }
}

\subsection{RQ2: How effective is \tool in distortion correction?}

The distortion correction task builds upon the results of distortion detection. We evaluate the effectiveness of correction by examining code lines marked with the ``\textit{//fixed}'' tag, which indicates that these lines have been successfully corrected. Lines classified as type I4 (redundant code) are removed during correction and therefore excluded from the statistics.

To provide a comprehensive evaluation, we compare \tool with two categories of baselines:

\begin{itemize}
    \item \textbf{Commercial off-the-shelf LLMs:} These include \promptd, \prompte, \promptz, and \degpt. They are general-purpose models typically used in zero-shot or prompt-based settings, without adaptation to the challenges of decompilation or binary analysis.

    \item \textbf{Domain-specific fine-tuned LLMs:} These include \textit{LLM4Decompile} and \textit{ReSym}, which have been specifically fine-tuned or developed to address binary analysis and code recovery, making them more suited for tasks involving the restoration of decompiled code to source code form.

\end{itemize}

\paragraph{Results with commercial off-the-shelf LLMs}
As shown in Table~\ref{tab:metrics}, \tool achieves the highest average fix rate (FR = 0.94) and corrected fix rate (CFR = 0.64) across all datasets. In contrast, \prompte and \promptd show competitive but slightly lower performance, while \degpt and \promptz exhibit clear limitations when handling more complex distortions. These results highlight the importance of incorporating distortion-type detection to guide LLM-based correction effectively.

\paragraph{Results with domain-specific fine-tuned LLMs}
Table~\ref{tab:baseline-comparison} presents the results on Ghidra-decompiled code. While both \textit{LLM4Decompile} and \textit{ReSym} demonstrate notable correction capabilities (with an average CFR of 0.44–0.45), their performance is consistently lower than that of \tool. Specifically, \textit{LLM4Decompile}, despite being fine-tuned, struggles with longer functions due to generation instability, whereas \textit{ReSym}, although effective in variable renaming, suffers from limited semantic coverage since it does not account for control-flow or syntactic distortions.

\noindent\colorbox{gray!20}{
    \parbox{\dimexpr\linewidth-2\fboxsep}{
    \textbf{Answer to RQ2:} From the experimental results, it is evident that \tool outperforms both (i) commercial off-the-shelf LLM approaches, which lack the capability to understand compiler-induced distortions, and (ii) domain-specific fine-tuned LLMs, which remain prone to instability or partial coverage. By explicitly modeling distortion types within a unified detection-correction pipeline, \tool strikes a robust balance between generality and practical performance.
    }
}

\begin{table*}[htbp]
\centering
\scriptsize
\caption{Performance Comparison of Different Approaches on Distortion Correction across Various Datasets with FR and CFR Metrics.}
\resizebox{\textwidth}{!}{ 
\begin{tabular}{|l|c|c|c|c|c|c|c|c|c|c|c|c|c|c|c|c|}
\hline
\textbf{Approach}  & \multicolumn{2}{c|}{\textbf{ImageMagick}} & \multicolumn{2}{c|}{\textbf{curl}} & \multicolumn{2}{c|}{\textbf{putty}} & \multicolumn{2}{c|}{\textbf{CAlgorithm}} & \multicolumn{2}{c|}{\textbf{coreutils}} & \multicolumn{2}{c|}{\textbf{OpenSSL}} & \multicolumn{2}{c|}{\textbf{SQLite}} & \multicolumn{2}{c|}{\textbf{Average}} \\ \hline
  \textbf{Metrics} & \textbf{FR} & \textbf{CFR} & \textbf{FR} & \textbf{CFR} & \textbf{FR} & \textbf{CFR} & \textbf{FR} & \textbf{CFR} & \textbf{FR} & \textbf{CFR} & \textbf{FR} & \textbf{CFR} & \textbf{FR} & \textbf{CFR} & \textbf{FR} & \textbf{CFR} \\ \hline
\textbf{\promptz}  & 0.8  & 0.13 & 0.8  & 0.17 & 0.73 & 0.17 & 0.8  & 0.21 & 0.77 & 0.18 & 0.77 & 0.17 & 0.6  & 0.15 & 0.75 & 0.17 \\ \hline
\textbf{\degpt}       & 0.8  & 0.29 & 0.85 & 0.3  & 0.87 & 0.4  & 0.93 & 0.48 & 0.85 & 0.39 & 0.8  & 0.38 & 0.73 & 0.33 & 0.83 & 0.37 \\ \hline
\textbf{\promptd} & 0.87 & 0.62 & 0.92 & 0.58 & 0.94 & 0.43 & 0.96 & 0.71 & 0.92 & 0.55 & 0.9  & 0.53 & 0.93 & 0.6  & 0.92 & 0.57 \\ \hline
\textbf{\prompte} & 0.89 & 0.56 & 0.91 & 0.53 & 0.91 & 0.44 & 0.96 & 0.64 & 0.91 & 0.49 & 0.91 & 0.54 & 0.9  & 0.58 & 0.91 & 0.54 \\ \hline
\textbf{\tool} & 0.92 & 0.67 & 0.96 & 0.61 & 0.95 & 0.61 & 0.96 & 0.76 & 0.92 & 0.63 & 0.92 & 0.61 & 0.96 & 0.6  & 0.94 & 0.64 \\ \hline
\end{tabular}
}
\label{tab:metrics}
\end{table*}

\begin{table*}[h]
\centering
\normalsize
\caption{Correction Performance Comparison of Domain-specific Fine-tuned LLMs.}
\label{tab:baseline-comparison}
\resizebox{\textwidth}{!}{
\begin{tabular}{|l|c|c|c|c|c|c|c|c|c|c|c|c|c|c|c|c|}
\hline
\textbf{Approach} & \multicolumn{2}{c|}{\textbf{ImageMagick}} & \multicolumn{2}{c|}{\textbf{curl}} & \multicolumn{2}{c|}{\textbf{putty}} & \multicolumn{2}{c|}{\textbf{CAlgorithm}} & \multicolumn{2}{c|}{\textbf{coreutils}} & \multicolumn{2}{c|}{\textbf{OpenSSL}} & \multicolumn{2}{c|}{\textbf{SQLite}} & \multicolumn{2}{c|}{\textbf{Average}} \\
\hline
\textbf{Metrics} & \textbf{FR} & \textbf{CFR} & \textbf{FR} & \textbf{CFR} & \textbf{FR} & \textbf{CFR} & \textbf{FR} & \textbf{CFR} & \textbf{FR} & \textbf{CFR} & \textbf{FR} & \textbf{CFR} & \textbf{FR} & \textbf{CFR} & \textbf{FR} & \textbf{CFR} \\
\hline
\textbf{LLM4Decompile} & 0.47 & 0.27 & 0.75 & 0.58 & 0.77 & 0.62 & 0.91 & 0.56 & 0.80 & 0.33 & 0.90 & 0.46 & 0.31 & 0.27 & \textbf{0.70} & \textbf{0.44} \\
\textbf{ReSym} & 0.81 & 0.46 & 0.85 & 0.43 & 0.91 & 0.43 & 0.89 & 0.49 & 0.87 & 0.51 & 0.79 & 0.41 & 0.74 & 0.41 & \textbf{0.84} & \textbf{0.45} \\
\hline
\end{tabular}
}
\end{table*}

\subsection{RQ3: How does the impact of each component of \tool on its overall effectiveness?}

In this section, we evaluate the effectiveness of the individual components of \tool, specifically the \textbf{Retrieval-Augmented Generation}, the \textbf{Variable Dependency Algorithm}, and the \textbf{Dynamic Semantic Intensity Retrieval Algorithm}. To achieve this, we conducted ablation studies to assess the performance and efficiency of each component across different levels of function complexity and datasets.


For the RAG component, we demonstrate its impact on the overall effectiveness of \tool. For the Variable Dependency Algorithm, we analyzed decompiled functions by sampling at regular 10-line intervals. This strategy enabled us to systematically examine the algorithm’s performance as function complexity increased.

For the Dynamic Semantic Intensity Retrieval Algorithm, we introduced two baseline approaches for comparison: randomly selecting code lines and retrieving all code lines. These comparisons provide insights into how well the proposed algorithm balances efficiency and effectiveness compared to more simplistic selection methods.

\subsubsection{Retrieval-Augmented Generation} 
\label{sec:Effectiveness of RAG}
In the distortion detection experiment(see \ref{sec:Distortion detection experiment}), we validated the effectiveness of RAG. The experimental results presented with RAG included the overall performance of \tool. Without RAG, the results would be identical to those obtained using \prompte and \promptd.

\subsubsection{Variable Dependency Algorithm} 
\label{sec:Effectiveness of Algorithm}
To evaluate the efficiency of the variable dependency algorithm across different ranges of decompiled function lines, we conducted a detailed analysis focusing on functions with at least 30 lines. As shown in Table~\ref{tab:vad-performance}, the recall of redundant variables and the average processing time are reported for different line intervals. Notably, for functions exceeding 90 lines, the recall drops to zero, indicating that the algorithm fails to identify redundant variables in highly complex cases.

For functions with more than 30 lines, the recall starts at 0.31 and gradually decreases as the function length increases, eventually reaching 0 for functions with over 80 lines. This indicates a significant drop in performance as function complexity increases. In terms of processing time, functions with over 30 lines take 11.1 seconds on average, and the time slightly decreases to 6 seconds for functions with over 80 lines.

\begin{figure}[htbp]
    \centering
    \includegraphics[width=\linewidth, height=0.1\textheight]{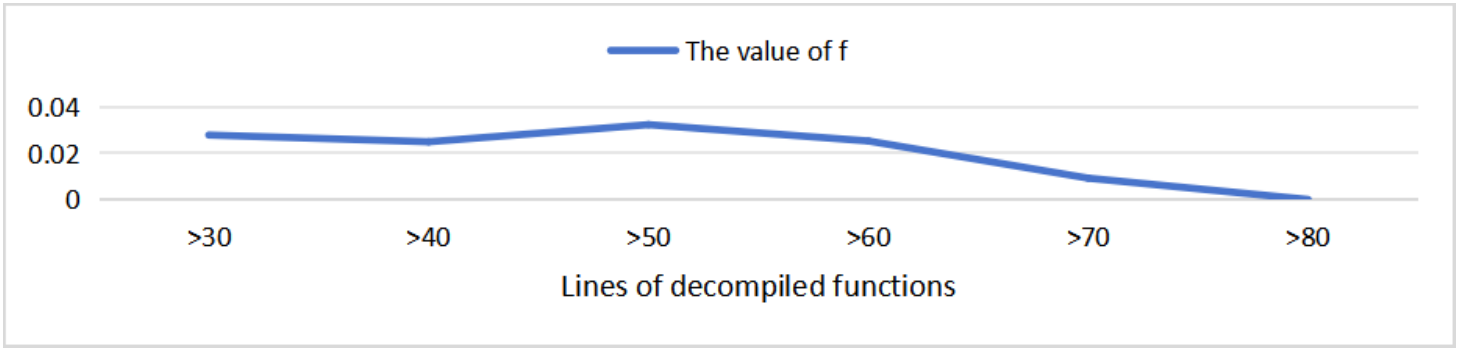}
    \caption{The value of f across different decompiled function line ranges}
    \label{fig:f}
\end{figure}

To assess the algorithm’s efficiency, we aim for a high recall with lower processing time. We define \textbf{recall} as the proportion of correctly identified distorted lines among all true distorted lines in a function. Based on this, we introduce a correlation factor $f$, defined as $f = \frac{\text{Recall}}{\text{Time}}$. A larger $f$ value signifies better algorithmic performance. As depicted in Fig.~\ref{fig:f}, the highest $f$ value is achieved at 50 lines, indicating that 50 lines are the optimal balance between efficiency and accuracy when processing decompiled functions. Therefore, we consider functions with more than 50 lines as \textbf{long functions}, which require chunk-based processing.

\begin{table}[htbp]
\footnotesize
\centering
\caption{Performance of Redundant Variable Recall and Time Across Decompiled Function Lines}
\begin{tabular}{|c|c|c|c|c|c|c|}
\hline
\textbf{Lines} & \textbf{30} & \textbf{40} & \textbf{50} & \textbf{60} & \textbf{70} & \textbf{80} \\ \hline
\textbf{Recall} & 0.31 & 0.23 & 0.24 & 0.18 & 0.06 & 0.00 \\ \hline
\textbf{Time (s)} & 11.1 & 9.2 & 7.4 & 7.1 & 6.5 & 6.0 \\ \hline
\end{tabular}
\label{tab:vad-performance}
\end{table}

\subsubsection{Dynamic Semantic Intensity Retrieval Algorithm} 

To evaluate the efficiency of the Dynamic Semantic Intensity Retrieval Algorithm, we adopted two baseline methods: randomly selecting six code lines and retrieving all code lines. We compared the performance of these three methods in terms of processing time, average number of query tokens, accuracy of distortion detection, and precision.

Table~\ref{tab:dsir-performance} indicates that the Dynamic Semantic Intensity Retrieval Algorithm has a slightly higher processing time and token count than the random selection method but is significantly more efficient than the method that retrieves all code lines. Specifically, \tool achieved a processing time of 4.51 seconds and a token count of 1,502.23, demonstrating its effectiveness in balancing efficiency and performance. In contrast, the random selection method had a slightly shorter processing time of 4.42 seconds and a token count of 1,402.08, but performed worse in terms of accuracy (0.81) and precision (0.84). This is mainly because the random selection of code lines does not guarantee the inclusion of the most representative code segments.

On the other hand, the all code lines retrieval method, while achieving relatively high accuracy (0.86) and precision (0.86), resulted in a significantly longer processing time (7.94 s) and a higher token count (1,641.55). This method's drawback lies in its inclusion of too many irrelevant or distorted code lines, adversely affecting the reasoning capabilities of LLMs and reducing overall efficiency.

The Dynamic Semantic Intensity Retrieval Algorithm achieved the highest accuracy (0.91) and precision (0.88). This suggests that by dynamically selecting code lines based on the semantic intensity and size of the input function, the method effectively filters out irrelevant or detrimental information, enhancing distortion detection performance. This strategy ensures the quality of the model's reasoning while avoiding unnecessary computational overhead, thereby optimizing retrieval efficiency and maintaining high performance.

\begin{table}[htbp]
\centering
\footnotesize
\caption{Performance of Dynamic Semantic Intensity Retrieval Algorithm}
\begin{tabular}{|l|c|c|c|c|}
    \hline
    \textbf{Approach} & \multicolumn{2}{c|}{\textbf{Performance}} & \multicolumn{2}{c|}{\textbf{Metrics}} \\ \cline{2-5}
    & \textbf{Time (s)} & \textbf{Tokens} & \textbf{Accuracy} & \textbf{Precision} \\ \hline
    \tool & 4.51 & 1,502.23 & 0.91 & 0.88 \\ \hline
    Random code & 4.42 & 1,402.08 & 0.81 & 0.84 \\ \hline
    All code & 7.94 & 1,641.55 & 0.86 & 0.86 \\ \hline
\end{tabular}
\label{tab:dsir-performance}
\end{table}

\begin{table}[htbp]
\centering
\footnotesize
\caption{Comparison of Average Token Count and Processing Time across Different Approaches for Analyzing Decompiled Functions}
\begin{tabular}{|l|r|r|}
    \hline
    \textbf{Approach} & \textbf{Tokens} & \textbf{Time (s)} \\ \hline
    \promptz & 1,424.3 & 3.2 \\ \hline
    \promptd & 2,836.4 & 5.66 \\ \hline
    \prompte & 4,154.2 & 6.08 \\ \hline
    \tool & 2,982.5 & 8.99 \\ \hline
    \degpt & 4,966.8 & 11.09 \\ \hline
\end{tabular}
\label{tab:performance_metrics}
\end{table}

\noindent\colorbox{gray!20}{
    \parbox{\dimexpr\linewidth-2\fboxsep}{
    \textbf{Answer to RQ3:} The ablation study demonstrates that RAG plays a crucial role within \tool. The Dynamic Semantic Intensity Retrieval Algorithm effectively selects the decompiled code lines with the highest semantic intensity, striking a balance between real-world performance and token usage. The Variable Dependency Algorithm shows significant advantages in handling redundant code, and using 50 lines as the threshold offers a reasonable trade-off between recall and processing time.
    }
}

\subsection{RQ4: How is the efficiency of \tool?}
Table~\ref{tab:performance_metrics} presents the average token count and processing time required to analyze a decompiled function. Firstly, \promptz shows the smallest number of tokens and the shortest processing time, as expected. It uses only basic prompts and directly outputs the results. \promptd follows it with efficiency slightly reduced due to the inclusion of additional fidelity-related prompts and the dual tasks of distortion detection and correction. Its token count is nearly doubled compared to \promptz's. \prompte exhibits a significant increase in both token count and processing time, as it involves three examples. This additional context contributes to a higher token count and longer processing duration. \tool achieves a balanced performance, with a modest increase in token count compared to \promptd. Since it includes only a small set of similar distortion codes retrieved from the distortion database as context, the token count does not increase excessively compared to \promptd's. \degpt has the highest token count and processing time, which aligns with its method. It involves three roles and requires multiple checks to generate the final result, leading to relatively lower efficiency.

\noindent\colorbox{gray!20}{
    \parbox{\dimexpr\linewidth-2\fboxsep}{
    \textbf{Answer to RQ4:} The experimental results demonstrate that \tool achieves favorable outcomes in terms of both token usage and processing time. This suggests that \tool strikes an effective balance between efficiency and performance.
    }
}

\begin{table*}[h]
\centering
\normalsize
\caption{Detection and correction performance of \tool across different decompiler and model settings.}
\label{tab:cross-backend}
\renewcommand{\arraystretch}{1.15}
\resizebox{\textwidth}{!}{
\begin{tabular}{|c|l|cc|cc|cc|cc|cc|cc|cc|c|}
\hline
\textbf{Setting} & \textbf{Metric} & \multicolumn{2}{c|}{\textbf{ImageMagick}} & \multicolumn{2}{c|}{\textbf{curl}} & \multicolumn{2}{c|}{\textbf{putty}} & \multicolumn{2}{c|}{\textbf{CAlgorithm}} & \multicolumn{2}{c|}{\textbf{coreutils}} & \multicolumn{2}{c|}{\textbf{OpenSSL}} & \multicolumn{2}{c|}{\textbf{SQLite}} & \textbf{Avg} \\
\hline
\multirow{2}{*}{\shortstack[c]{\tool(GPT-4o,\\IDA, T=0)}} 
  & Acc / Pr     & 0.91 & 0.78 & 0.90 & 0.79 & 0.88 & 0.84 & 0.91 & 0.76 & 0.86 & 0.76 & 0.89 & 0.78 & 0.88 & 0.89 & \textbf{0.89 / 0.80} \\
  & FR / CFR     & 0.97 & 0.68 & 0.97 & 0.72 & 0.97 & 0.62 & 0.98 & 0.65 & 0.97 & 0.65 & 0.97 & 0.72 & 0.96 & 0.62 & \textbf{0.97 / 0.67} \\
\hline
\multirow{2}{*}{\shortstack[c]{\tool(GPT-4o,\\Ghidra, T=0)}}
  & Acc / Pr     & 0.89 & 0.71 & 0.89 & 0.75 & 0.85 & 0.81 & 0.90 & 0.79 & 0.91 & 0.71 & 0.88 & 0.73 & 0.87 & 0.78 & \textbf{0.88 / 0.75} \\
  & FR / CFR     & 0.94 & 0.65 & 0.92 & 0.68 & 0.89 & 0.63 & 0.91 & 0.75 & 0.96 & 0.56 & 0.92 & 0.57 & 0.93 & 0.62 & \textbf{0.92 / 0.64} \\
\hline
\multirow{2}{*}{\shortstack[c]{\tool(DeepSeek-chat,\\IDA, T=0)}}
  & Acc / Pr     & 0.91 & 0.70 & 0.89 & 0.73 & 0.91 & 0.78 & 0.90 & 0.74 & 0.89 & 0.73 & 0.91 & 0.73 & 0.87 & 0.76 & \textbf{0.90 / 0.74} \\
  & FR / CFR     & 0.96 & 0.79 & 0.96 & 0.77 & 0.97 & 0.52 & 0.95 & 0.75 & 0.95 & 0.71 & 0.96 & 0.78 & 0.99 & 0.69 & \textbf{0.96 / 0.72} \\
\hline
\end{tabular}
}
\end{table*}

\subsection{RQ5: How robust and generalizable is \tool across compilers, decompilers, and LLM backends?}

To assess whether \tool maintains stable performance under different compilation, decompilation, and model settings, we further evaluate its robustness and generalizability.

\subsubsection{Evaluation Across Compiler Optimization Levels}
We examine the impact of compiler optimization levels (\texttt{-O1}, \texttt{-O2}, \texttt{-O3}) on distortion detection and correction. As shown in Table~\ref{tab:opt-level-singlecol}, \tool achieves consistently high performance across all three settings. Detection accuracy remains stable at around 0.89–0.90, with precision gradually decreasing from 0.78 at \texttt{-O1} to 0.70 at \texttt{-O3}. For correction, the fix rate (FR) stays above 0.96 across all levels, while the corrected fix rate (CFR) shows a slight decline from 0.60 to 0.59 as optimizations become more aggressive. These results indicate that, although aggressive compiler optimizations introduce additional distortions (particularly reducing precision), \tool maintains strong robustness without significant overall degradation. Note that the performance under \texttt{-O0} was reported separately in Table~\ref{tab:detection} and Table~\ref{tab:metrics}, where similarly high accuracy and fix rates were achieved.

This small performance gap can be attributed to two factors. First, our definition of six distortion types and the construction of a comprehensive decompiled distortion database ensure that both unoptimized distortions (\texttt{-O0}) and compiler-induced transformations (\texttt{-O1}, \texttt{-O2}, \texttt{-O3}) are well covered. Combined with RAG-based retrieval, \tool consistently supplies the LLM with semantically similar distortion exemplars, making it less sensitive to optimization levels. Second, this finding is consistent with prior observations~\cite{li2025sv} that LLMs are overly reliant on pattern matching rather than deep logical reasoning for complex code. As long as surface distortion patterns are retrievable from the database, \tool can effectively guide recovery, thus narrowing the performance differences across optimization levels.

\begin{table}[h]
\centering
\scriptsize
\caption{Detection and correction performance of \tool under different compiler optimization levels.}
\begin{tabular}{|l|c|c|c|c|c|c|}
\hline
\textbf{Approach} & \multicolumn{2}{c|}{\textbf{-O1}} & \multicolumn{2}{c|}{\textbf{-O2}} & \multicolumn{2}{c|}{\textbf{-O3}} \\
\hline
\textbf{Detection Metrics} & \textbf{Acc} & \textbf{Pr} & \textbf{Acc} & \textbf{Pr} & \textbf{Acc} & \textbf{Pr} \\
\hline
\tool (Detection) & 0.90 & 0.78 & 0.89 & 0.75 & 0.89 & 0.70 \\
\hline
\textbf{Correction Metrics} & \textbf{FR} & \textbf{CFR} & \textbf{FR} & \textbf{CFR} & \textbf{FR} & \textbf{CFR} \\
\hline
\tool (Correction) & 0.96 & 0.60 & 0.97 & 0.61 & 0.96 & 0.59 \\
\hline
\end{tabular}
\label{tab:opt-level-singlecol}
\end{table}

\subsubsection{Generalizability Across Decompilers and Model Backends}
To test the portability of our approach across different toolchains, we vary both the decompiler (IDA Pro vs. Ghidra) and the underlying LLM (GPT-4o vs. DeepSeek-chat). The results in Table~\ref{tab:cross-backend} show that \tool achieves stable performance across all settings.  
\tool(GPT-4o, Ghidra, T=0) yields only marginally lower accuracy and CFR compared with the IDA-based setup, confirming its \textbf{decompiler-agnostic} capability.  
Meanwhile, \tool(DeepSeek-chat, IDA, T=0) delivers comparable detection quality and even higher CFR (0.72 on average), validating the \textbf{model-agnostic} property of our design.  

\noindent\colorbox{gray!20}{
    \parbox{\dimexpr\linewidth-2\fboxsep}{
    \textbf{Answer to RQ5:} \tool demonstrates strong robustness and generalizability across compiler optimization levels, decompilers, and LLM backends. This robustness is primarily enabled by the distortion database and RAG-based guidance, which provide LLMs with representative distortion exemplars and mitigate their reliance on deeper logical reasoning. As a result, our method can be reliably integrated into diverse reverse engineering pipelines with minimal performance degradation.
    }
}

\section{Discussion}
\label{sec:Discussion}

\subsection{Manual Evaluation}

Manual evaluation was conducted by seven team members and served as a critical component of our dataset construction and verification pipeline. Specifically, it involved aligning source and decompiled code, identifying and labeling distortions, and validating the corresponding repair outcomes. This annotation process formed the basis of our 620-pair ground truth dataset used for evaluation. The participants included two researchers, two PhD students, and three master's students with relevant expertise in software engineering and binary analysis. They were divided into two groups: one member performed the initial annotation, while the other conducted independent verification. Discrepancies were resolved through discussion or literature reference to ensure consensus and annotation reliability.

\subsection{Threats to Validity}




Despite our efforts to ensure robustness, several threats to validity remain.

\textbf{LLM Randomness.} Large language models inherently exhibit non-deterministic behavior: identical inputs can yield different outputs across runs. This variability may affect the reproducibility and stability of our results. To mitigate this, we conducted multiple trials and reported average performance. Moreover, we provide results generated at zero temperature, which significantly reduces randomness and enhances reproducibility. Nonetheless, some degree of stochasticity remains intrinsic to current LLMs and should be considered when interpreting outcomes.

\textbf{Potential Data Leakage.} Another concern is corpus leakage—where pretraining data may overlap with source code used in our experiments. While we observed inconsistent outputs even for repeated inputs, indicating the absence of direct memorization, this cannot fully rule out indirect exposure. We repeated all experiments multiple times to average out such effects, but further analysis is needed to fully assess this risk.

\subsection{User Study}
\label{sec: user study}

The user study was designed to assess the effectiveness of our repair method in assisting real-world reverse engineering tasks. We recruited 15 participants outside the development team, all with prior experience in binary analysis or software reverse engineering. Each participant was asked to compare raw decompiled code and code repaired by our method, and then evaluate them based on readability, correctness, and ease of semantic recovery. The study was conducted in a controlled setting with standardized tasks, and detailed instructions were provided to minimize bias. Full procedures and aggregated results are provided in Appendix~\ref{sec:Supplementary User Study Details}.

\subsection{Limitations and Future Work}


\textbf{Scalability to Complex Functions.}  
\tool handles long functions differently in the detection and correction phases. For detection, functions exceeding 50 lines are divided into smaller chunks, allowing the model to maintain accuracy on large inputs. In contrast, correction operates on the entire function to preserve variable naming consistency and global semantics—segmenting at this stage would disrupt cross-chunk dependencies and reduce fix quality. As function size increases beyond the LLM’s effective reasoning scope, correction performance may degrade due to limited long-range dependency modeling. We note this as a general limitation of current LLM-based repair systems. Future improvements in model architecture or techniques like hierarchical generation may help enhance scalability.

\textbf{Coverage of Distortion Types.}  
Our distortion database is manually built on top of functions sampled from the \texttt{Dramko} dataset. While this offers solid coverage of commonly encountered distortions, it may fall short in representing rarer or more complex semantic anomalies. In such cases, distortions must be manually identified and incorporated by experts. We aim to automate this discovery process in future work by combining program differencing techniques with anomaly detection methods, thereby enabling more complete and scalable construction of distortion benchmarks.

\textbf{Downstream Tasks.} Our current research focuses on automated detection and correction of decompilation distortions. Future studies will investigate the impact of these distortions on downstream tasks, such as binary similarity detection, to enhance the reliability of analyses dependent on decompiled code.

\textbf{Semantic Consistency Verification.} Although \tool corrects decompilation distortions, the repaired code often cannot be recompiled and executed to verify semantic consistency due to missing type information or incomplete constructs. Future work will develop methods to generate recompilable code, such as inferring types and reconstructing function signatures, and validate functionality through execution, improving the reliability of semantic equivalence assessments.

\section{Related Work}
\label{sec:RelatedWork}

\subsection{Fidelity in Decompilation}

Decompilation presents key challenges such as variable name recovery, type inference, and code structure reconstruction. While machine learning (ML) and large language models (LLMs) have advanced in these areas, limitations remain.
ML techniques have been applied to enhance decompilation outputs, particularly in variable name recovery. The \textit{DIRE} model~\cite{dramko2023dire} uses probabilistic methods to leverage both lexical and structural information for this task. However, it struggles with generalization. To address this, \textit{VARBERT}~\cite{pal2024len}, based on BERT, employs transfer learning to improve variable name prediction, setting new benchmarks.
LLMs have made significant strides in decompilation. \textit{LLM4Decompile}~\cite{tan2024llm4decompile} introduces a range of models trained specifically for decompilation tasks. \textit{ReSym}~\cite{xie2024resym} combines LLMs with program analysis to recover variable names and types, improving overall accuracy.

\subsection{Prompt Engineering}

Prompt engineering focuses on designing prompts to enhance LLM performance. Recent work emphasizes prompt design, contextual prompting, and task-specific tuning.
Early studies show that prompt phrasing affects LLM output. Webson's~\cite{webson2021prompt} explores the effectiveness of prompt design, revealing that models often perform well even with irrelevant prompts, raising questions about prompt understanding. Cao's~\cite{cao2023study} investigates how prompt templates impact debugging performance in ChatGPT.
Contextual prompting incorporates task-specific context. Wei's~\cite{wei2021finetuned} improves zero-shot learning by finetuning LLMs using instruction templates, leading to better performance on unseen tasks.

\subsection{Retrieval-Augmented Generation (RAG)}
RAG enhances LLMs by retrieving external knowledge, which improves consistency and reduces hallucinations in generated content.
RAG stabilizes output by incorporating external knowledge, ensuring consistent results across similar inputs~\cite{cuconasu2024power,guo2024ft2ra}.
By grounding outputs in verified information, RAG reduces the likelihood of hallucinations, leading to more accurate responses~\cite{wu2023ragtruth,li2024enhancing}.
\section{Conclusion}
\label{sec:Conclusion}

In this study, we addressed decompilation distortions in the closed-source context using large language models (LLMs) and the Retrieval-Augmented Generation (RAG) technique. Our framework, \tool, effectively detected and corrected six types of distortions, achieving 89\% accuracy, 83\% precision, a 94\% fix rate, and a 64\% correction rate. These results significantly enhance decompiled code readability and accuracy, offering a valuable tool for reverse engineering.

\section{Data Availability}
\label{sec:ata Availability}
Our source code and dataset are available at \cite{fidelitygpt}.



\section*{Acknowledgment}

This work was supported by the National Natural Science Foundation of China (Key Program, Grant No. 62332005).

\bibliographystyle{IEEEtran}
\bibliography{reference}

\appendices

\section{Study and Dataset Overview}
\label{sec:Appendix}

\subsection{Case Studies}

To empirically validate the challenges in decompilation and the limitations of existing approaches, we present three concrete case studies analyzing common distortions in closed-source binary decompilation. These studies demonstrate how decompiler outputs diverge from original source code, the impact of retrieval-augmented generation (RAG) in distortion detection, and the cascading effects of false positives/negatives during correction. Through these analyses, we highlight why current taxonomies and manual inspection methods fail to address practical decompilation scenarios where source references are unavailable.

\subsubsection{Limitations of Existing Taxonomy in Closed-Source Scenarios} 
\label{sec:Case study I}
\begin{figure*}[htbp]
    \centering
    \includegraphics[width=\textwidth]{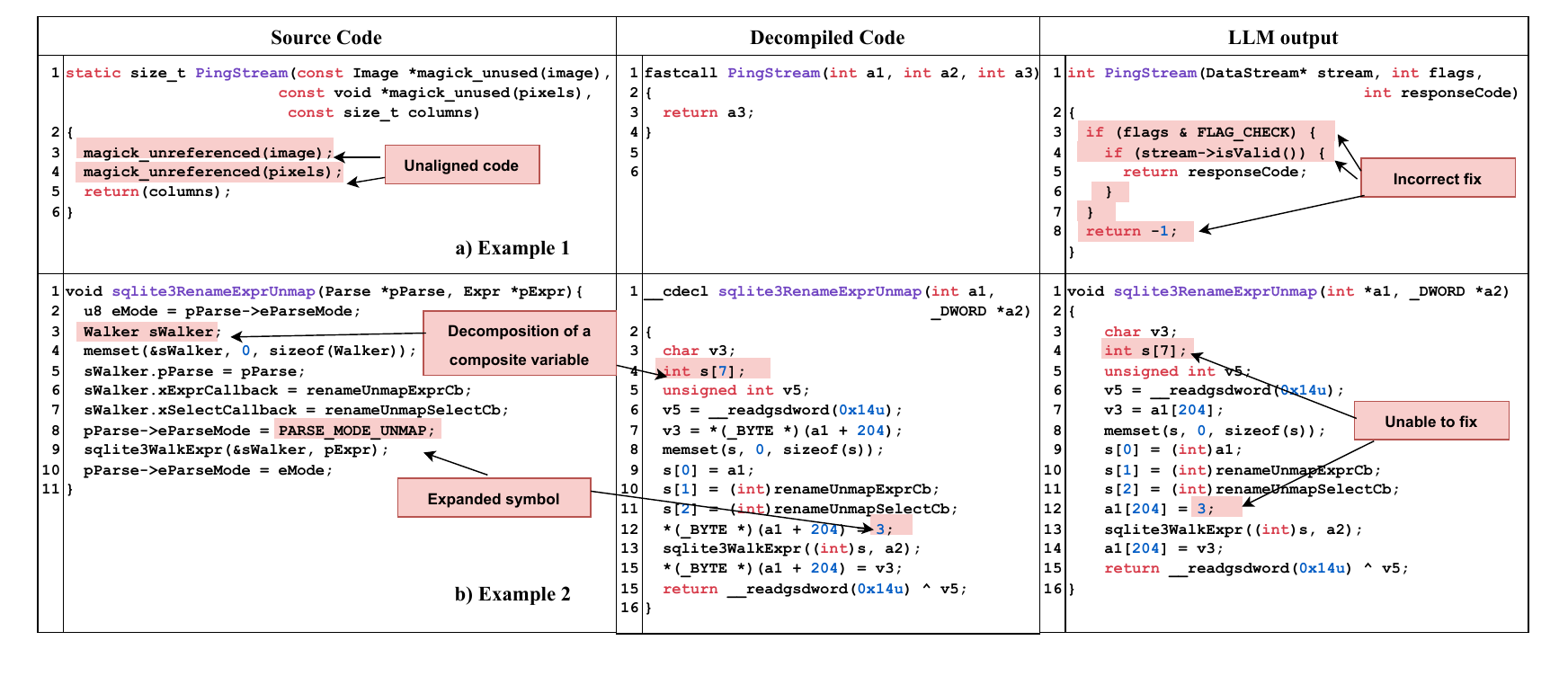}
    \caption{Examples of Taxonomy-Defined Decompilation Distortions Intractable in Closed-Source Environments}
    \label{fig:motivation0}
\end{figure*}

To illustrate these challenges and the limitations of existing taxonomy, we present Figure~\ref{fig:motivation0}, which showcases specific examples of decompilation distortions prevalent in closed-source scenarios. These examples, rooted in concepts from Dramko et al.'s taxonomy~\cite{dramko2024taxonomy}, include \textbf{unaligned code}, \textbf{decomposition of composite variables}, and \textbf{expanded symbols}. The figure juxtaposes the \textbf{source code} (the original, unavailable in closed-source settings), the \textbf{decompiled code} (the decompiler's output), and the \textbf{LLM output} (results from attempting correction with a large language model). These examples demonstrate why Dramko et al.'s taxonomy, while insightful, falls short in addressing practical decompilation issues when source code is absent.

\begin{enumerate}
    \item \textbf{Unaligned code} (Panel a): During compilation, unused code segments may be optimized out, leading to misalignment between the decompiled output and the original source. Without access to the source code, reconstructing these missing sections is challenging. As shown in the \textbf{LLM output} column, attempts to regenerate the code using an LLM often introduce semantic errors, deviating from the intended functionality.    
    \item \textbf{Decomposition of composite variables} and \textbf{expanded symbols} (Panel b): When functions directly utilize data structures, decompilers may split composite variables into individual elements, while user-defined macros are expanded into constants. Without the source code, accurately reconstructing the original variable composition or macro definitions is nearly impossible. The \textbf{LLM output} illustrates how such corrections often yield inaccurate or incomplete results, exceeding the capabilities of manual inspection.
\end{enumerate}

\subsubsection{Impact of RAG Retrieval on Distortion Detection} 
\label{sec:Case study II}
\begin{figure*}[htbp]
    \centering
    \includegraphics[width=\textwidth]{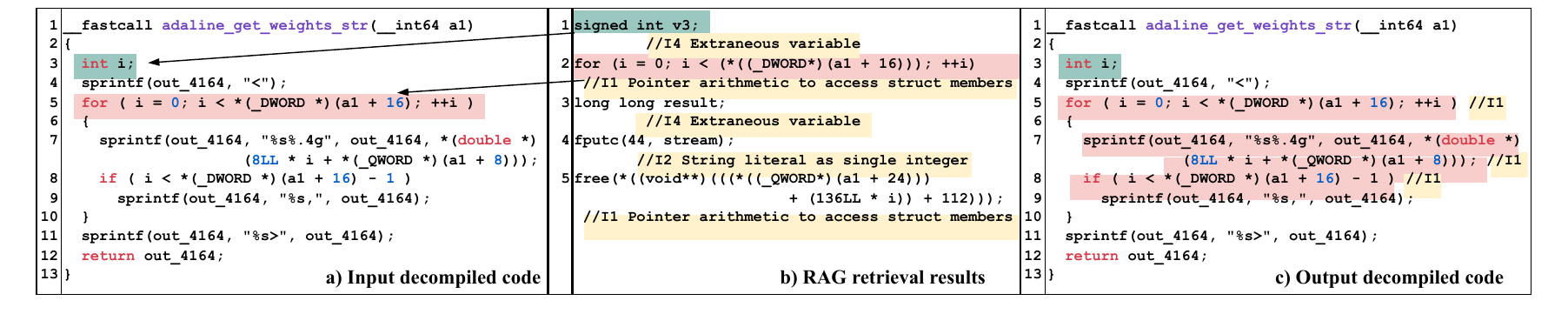}
    \caption{Impact of RAG Retrieval on Distortion Detection}
    \label{fig:Case study 1}
\end{figure*}

In this section, we discuss the impact of the RAG retrieval results on \tool. As depicted in Fig.~\ref{fig:Case study 1}, panel (a) shows the input decompiled code, while panel b illustrates the retrieval results from the decompilation distortion database, which are used as contextual prompts for the large language model. From panel (b), we can observe that the first retrieved line corresponds to a redundant variable, which is similar to line three of the decompiled code. However, the variable \textit{i} is not actually redundant, and the output does not result in a false positive. The second retrieved line is similar to the fifth line of the decompiled code and is correctly labeled as \textit{I1} in the output. The third, fourth, and fifth retrieved lines do not have any particularly similar counterparts in the decompiled code, and as input prompts, they do not affect the distortion detection results. Thanks to the well-defined prompt templates and the large language model's understanding of code, the output results, as shown in panel (c), demonstrate that the decompiled code detection is accurate.

\subsubsection{Effects of Detection Errors on Correction Outcomes} 
\label{sec:Case study III}
\begin{figure*}[htbp]
    \centering
    \includegraphics[width=\textwidth]{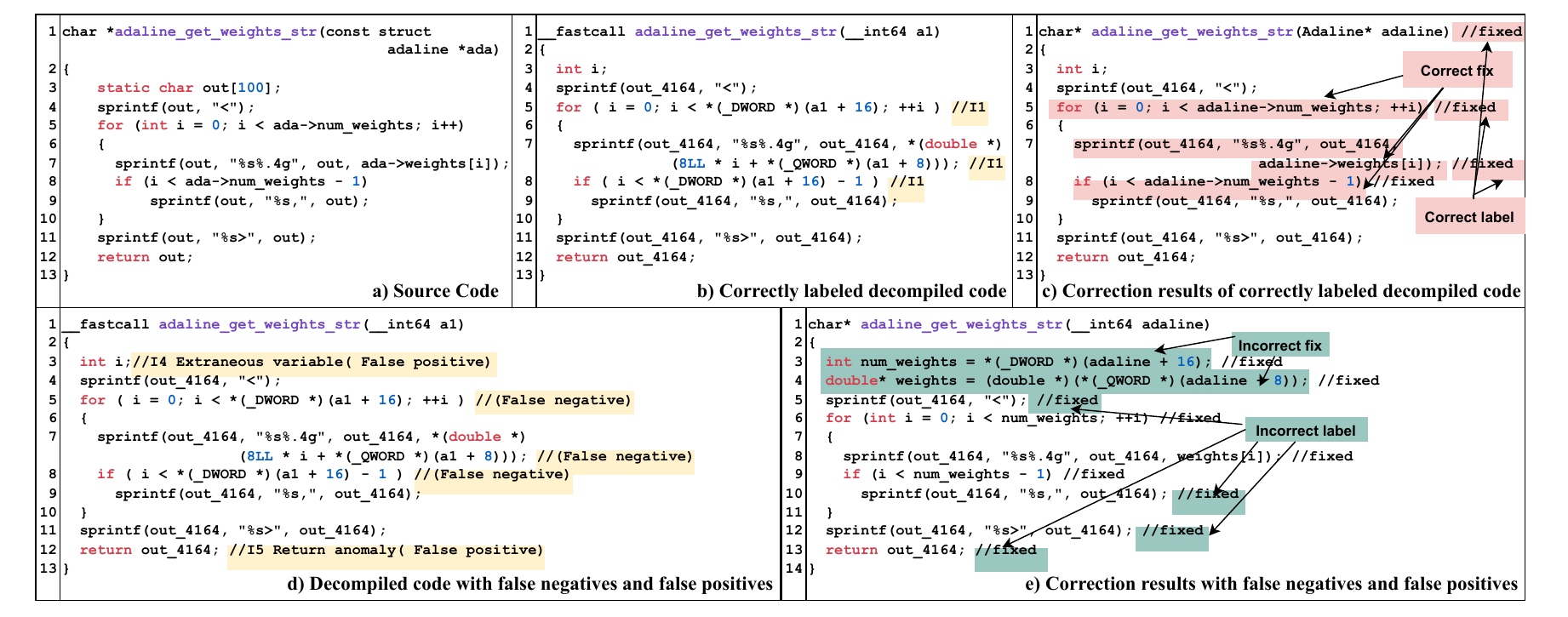}
    \caption{Effects of Detection Errors on Correction Outcomes}
    \label{fig:Case study 2}
\end{figure*}

In this section, we conduct a case study to examine the impact of false negatives and false positives during the distortion detection phase on the distortion correction phase, as shown in figure~\ref{fig:Case study 2}. Panel (a) presents the source code, while panels (b) and (d) depict the decompiled code with correct labels and the decompiled code with false negatives and false positives, respectively. Panel (c) shows the output for the decompiled code with correct labels. When compared to the source code, the correction phase effectively addresses the distortion issues, and the result closely resembles the original source code.

In panel (d), the third and twelfth lines are false positives, while the fifth, seventh, and eighth lines are false negatives. The correction results are shown in panel (e). For false negatives, the corrected code retains its original structure. However, for false positives, the code semantics may be altered, and incorrect "fixed" labels may be introduced, as seen in line 13.

It is important to note that in the source code shown in panel (a), the "out" variable on the third line is a static array. Due to compiler optimizations, the specific type could not be recovered during the decompilation process. As mentioned earlier (see \ref{sec:Distortion Issues in a More Realistic Context}), in the absence of source code references, this issue goes beyond the scope of code optimization. Therefore, this underscores why we place greater emphasis on false positives during the detection phase.

These case studies collectively reveal three critical insights:

\begin{enumerate}
    \item Source absence is irreplaceable: Even with LLMs, reconstructing optimized-out code or composite variables remains error-prone without source references.
    \item Context matters: Retrieval-augmented detection improves accuracy but depends on the semantic alignment of prompts and decompiled code.
    \item False positives are high-risk: False positives during detection propagate irreversible semantic errors in correction, whereas false negatives merely preserve decompiler output.
\end{enumerate}

\begin{table*}[ht]
\footnotesize
\centering
\caption{Comparison of statistics between evaluation subset and full dataset}
\label{tab:dataset_stats}
\resizebox{\textwidth}{!}{%
\begin{tabular}{l|cccc|cccc|cc}
\toprule
\textbf{Dataset} & \multicolumn{4}{c|}{\textbf{Lines per Function}} & \multicolumn{4}{c|}{\textbf{Tokens per Function}} & \multicolumn{2}{c}{\textbf{Large Functions}} \\
 & Max & Min & Median & Mean & Max & Min & Median & Mean & $>$50 lines & $>$1024 tokens \\
\midrule
\textbf{Evaluation Subset (620 functions)} & \textbf{503} & \textbf{4} & \textbf{20} & \textbf{30.1} & \textbf{4105} & \textbf{10} & \textbf{139} & \textbf{212.9} & \textbf{75 (12.1\%)} & \textbf{12 (1.9\%)} \\
\textbf{Full Dataset (46,941 functions)} & \textbf{3342} & \textbf{3} & \textbf{10.0} & \textbf{29.5} & \textbf{12538} & \textbf{8} & \textbf{33.0} & \textbf{110.1} & \textbf{5463 (11.6\%)} & \textbf{675 (1.4\%)} \\
\bottomrule
\end{tabular}
}
\end{table*}

\subsection{Supplementary User Study Details}
\label{sec:Supplementary User Study Details}
This appendix provides additional details for the user study evaluating \tool’s repaired decompiled code outputs, as discussed in Section~\ref{sec: user study}. The survey was distributed to participants with varying expertise in reverse engineering, who rated the repaired code’s readability, conciseness, accuracy, and semantic fidelity on a scale from 1 (Strongly Disagree) to 10 (Strongly Agree). Table~\ref{tab:user_survey} presents the survey instrument.

\begin{table}[htbp]
\centering
\caption{User Survey Instrument for Evaluating \tool’s Repaired Decompiler Output}
\label{tab:user_survey}
\begin{tabular}{|p{0.95\columnwidth}|}
\hline
\textbf{Introduction} \\
The survey evaluates the effectiveness of \tool, a framework for repairing decompiled code outputs. Participants reviewed example code pairs and rated agreement with statements on a scale: 1 (Strongly Disagree) to 10 (Strongly Agree). \\
\hline
\textbf{Participant Expertise} \\
Participants selected one: \\
- Basic: Limited experience in reading and understanding decompiled code. \\
- Intermediate: Moderate experience in working with decompiled code. \\
- Professional: Extensive experience in reverse engineering and decompiled code analysis. \\
\hline
\textbf{Questions} \\
1. The repaired decompiler output is easier to read and understand, particularly in terms of variable names, types, and dereferencing (e.g., accessing structure members through pointers or arrays), compared to the original output. \\
Rating Scale: [1 -- 2 -- 3 -- 4 -- 5 -- 6 -- 7 -- 8 -- 9 -- 10] \\
2. The repaired decompiler output reduces redundant code (e.g., unnecessary variables, meaningless assignments) and improves code conciseness and clarity, enhancing overall readability. \\
Rating Scale: [1 -- 2 -- 3 -- 4 -- 5 -- 6 -- 7 -- 8 -- 9 -- 10] \\
3. The repaired decompiler output is more accurate in representing the expected program structure (e.g., handling unexpected returns or type mismatches) compared to the original output. \\
Rating Scale: [1 -- 2 -- 3 -- 4 -- 5 -- 6 -- 7 -- 8 -- 9 -- 10] \\
4. The repaired decompiler output better preserves the intended functionality and semantics of the original source code compared to the original decompiler output. \\
Rating Scale: [1 -- 2 -- 3 -- 4 -- 5 -- 6 -- 7 -- 8 -- 9 -- 10] \\
5. Overall, the repaired decompiler output is more helpful in understanding and potentially debugging the decompiled code compared to the original decompiler output. \\
Rating Scale: [1 -- 2 -- 3 -- 4 -- 5 -- 6 -- 7 -- 8 -- 9 -- 10] \\
\hline
\textbf{Additional Feedback} \\
Participants were invited to provide optional comments or suggestions regarding the repaired decompiler output. \\
\hline

\end{tabular}
\end{table}

\begin{figure}[htbp]
    \centering
    \includegraphics[width=\linewidth, height=0.1\textheight]{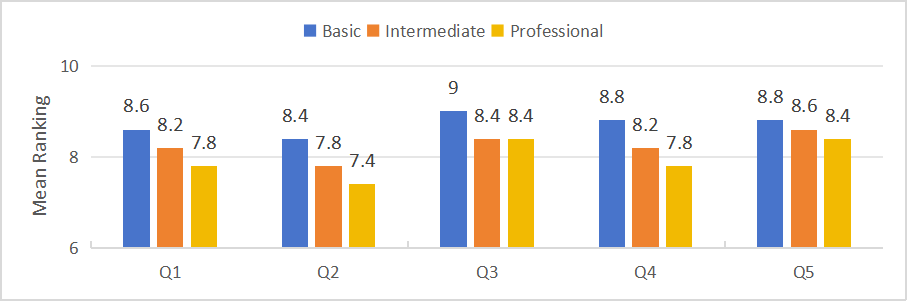}
    \caption{Mean user survey ratings for \tool’s repaired decompiler output}
    \label{fig:survey_ratings}
\end{figure}

Figure~\ref{fig:survey_ratings} shows mean survey ratings for \tool’s repaired outputs across 15 participants: 5 Basic (\textless{}1 year experience), 5 Intermediate (1--3 years), and 5 Professional (\textgreater{}3 years), with the survey instrument in Table~\ref{tab:user_survey}.


\subsection{Dataset Statistics and Representativeness}
\label{sec:dataset_analysis_appendix}

To validate the representativeness of the 620 annotated function pairs, we compared their structural properties with those of the full decompiled function dataset (46,941 functions in total). As shown in Table~\ref{tab:dataset_stats}, the evaluation subset exhibits slightly larger average function length (30.1 vs. 29.5 lines) and token count (212.9 vs. 110.1), indicating higher complexity.

While the subset's maximum function length (503 lines) and token count (4105 tokens) are lower than the dataset's global extremes (3342 lines, 12,538 tokens), these outliers are extremely rare—only 0.6\% and 0.1\% of functions exceed these respective thresholds. Therefore, the subset already covers over 99\% of practical real-world function cases.

These statistics confirm that the evaluation subset is structurally diverse and statistically representative.

\section{Artifact Appendix}

This artifact appendix provides a roadmap for setting up and evaluating the FidelityGPT artifact, as presented in the paper, ``FidelityGPT: Correcting Decompilation Distortions with Retrieval Augmented Generation.'' The artifact supports the detection and correction of distortions in decompiled functions using large language models (GPT-4o). It includes datasets, scripts, and configuration files to reproduce the paper's results.

\subsection{Description \& Requirements}

This section details the hardware, software, and dataset requirements necessary to recreate the experimental setup for the FidelityGPT artifact.

\subsubsection{How to access}
The artifact is hosted at \url{https://doi.org/10.5281/zenodo.17070171}.

\subsubsection{Hardware dependencies}
\begin{itemize}
  \item \textbf{CPU}: Intel Core i7-12700H or equivalent.
  \item \textbf{RAM}: 32 GB.
  \item \textbf{GPU}: NVIDIA RTX 3070Ti (8 GB VRAM) or equivalent.
  \item \textbf{Storage}: 2 GB free disk space.
\end{itemize}

\subsubsection{Software dependencies}
\begin{itemize}
  \item \textbf{Operating System}: Windows 11.
  \item \textbf{Python}: Version 3.9.
  \item \textbf{Dependencies}: Langchain 0.1.2, IDA Pro 7.5, and API libraries (e.g., \texttt{openai}) listed in \texttt{requirements.txt}.
  \item \textbf{APIs}: GPT-4o (requires API keys).
\end{itemize}

\subsubsection{Benchmarks}
The artifact includes the following datasets and scripts:
\begin{itemize}
  \item \textbf{Datasets}:
    \begin{itemize}
      \item \textbf{Ground truth functions}: Stored in \texttt{Ground truth/*.txt}, with functions separated by \texttt{/////}.
      \item \textbf{Decompiled functions}: Stored in \texttt{Dataset/*.txt}, with functions separated by \texttt{/////}.
      \item \textbf{Decompilation distortion database}: \texttt{fidelity\_new.c} (for IDA Pro) and \texttt{fidelity\_ghidra.c} (for Ghidra).
    \end{itemize}
  \item \textbf{Scripts}:
    \begin{itemize}
      \item \texttt{FidelityGPT.py}: Detects distortions in decompiled functions.
      \item \texttt{Correction.py}: Corrects distortions in decompiled functions.
      \item \texttt{PromptTemplate.py}: Provides all prompt templates, including those for variable dependency analysis, distortion detection, distortion correction, and baseline methods.
      \item \texttt{pattern\_matcher.py}: Implements the Dynamic Semantic Intensity Retrieval Algorithm, generating semantic strength weights from the decompilation distortion database, calculating semantic strength for input code lines, and extracting top-k lines for Retrieval-Augmented Generation (RAG).
      \item \texttt{variabledependcy.py}: Implements the Variable Dependency Algorithm, generating Program Dependence Graphs (PDGs) and extracting variable dependencies to determine redundancy.
      \item \texttt{Evaluation/Evaluation.py}: Evaluates detection results.
    \end{itemize}
\end{itemize}

\subsection{Artifact Installation \& Configuration}

To prepare the environment for evaluating the artifact:
\begin{enumerate}
  \item \textbf{Set up Python 3.9 virtual environment}:
\begin{lstlisting}[language=bash]
python -m venv venv
venv\Scripts\activate
\end{lstlisting}
  \item \textbf{Install dependencies}:
\begin{lstlisting}[language=bash]
pip install -r requirements.txt
\end{lstlisting}
\end{enumerate}

\subsection{Experiment Workflow}

The FidelityGPT artifact supports two experiments to validate the paper's claims:
\begin{itemize}
  \item \textbf{Detection Phase}: Uses \texttt{FidelityGPT.py} to detect distortions in decompiled functions, labeling them with distortion types (I1 to I6).
  \item \textbf{Correction Phase}: Uses \texttt{Correction.py} to correct distortions, producing functions labeled with \texttt{//fix}.
  \item \textbf{Evaluation Phase}: Uses \texttt{Evaluation/Evaluation.py} to evaluate detection results against ground truth data.
\end{itemize}

The workflow involves configuring \texttt{config.ini}, running detection and correction scripts, and evaluating results against ground truth data.

\subsection{Major Claims}

We have two major claims:
\begin{itemize}
  \item (C1): FidelityGPT effectively detects distortions in decompiled functions, achieving high accuracy and precision. This is proven by experiment (E1), with results reported in Table II of the paper.
  \item (C2): FidelityGPT effectively corrects distortions in decompiled functions, as measured by Fix Rate (FR) and Correct Fix Rate (CFR). This is proven by experiment (E2), with results reported in Table III of the paper.
\end{itemize}

\subsection{Configuration}

Before running the system, update the configuration file \texttt{config.ini}:

\begin{lstlisting}
[LLM]
model = gpt-4o
temperature = 0
api_key = sk-XXXX
api_base = XXXX 

[PATHS]
input_dir = Dataset_4_AE 
output_dir = Dataset_4_AE_output
knowledge_base = fidelity_new.c 
\end{lstlisting}

\begin{itemize}
  \item Input functions: \texttt{.txt} files, each with functions separated by \texttt{/////}.
  \item Distortion database: \texttt{fidelity\_new.c} (IDA Pro) or \texttt{fidelity\_ghidra.c} (Ghidra).
\end{itemize}

\subsection{Evaluation}

This section provides operational steps to validate the artifact's functionality and reproduce the paper's results. The experiments (E1 and E2) correspond to the major claims (C1 and C2).

\subsubsection{Experiment (E1): Detection Evaluation}

\textit{[Preparation]}
\begin{itemize}
  \item Create a dedicated folder for test functions and copy the inputs:
\begin{lstlisting}[language=bash]
mkdir Dataset_4_AE
cp testdata/*.txt Dataset_4_AE/
# alternatively:
cp Dataset/*.txt Dataset_4_AE/
\end{lstlisting}
\end{itemize}

\textit{[Execution]}
\begin{itemize}
  \item Run the detection script:
\begin{lstlisting}[language=bash]
python FidelityGPT.py
\end{lstlisting}
  \item Input: Files from \texttt{Dataset\_4\_AE/}. 
  \item Output: Results saved in \texttt{Dataset\_4\_AE\_output/}, where each function is labeled with distortion types (\texttt{I1--I6}) and separated by \texttt{/////}.
  \item For functions longer than 50 lines, the system applies chunk-based detection with a 5-line overlap. After detection:
  \begin{itemize}
    \item Manually merge chunked functions.
    \item Remove overlapping duplicate lines.
    \item Preserve the \texttt{/////} separator between functions.
  \end{itemize}
  \item Ensure line alignment before evaluation: each line in \texttt{model\_output.txt} must correspond exactly to the same function segment in \texttt{ground\_truth.txt}. In practice, we place both files side-by-side (e.g., in Excel) to verify alignment.
  \item Once aligned, run the evaluation script:
\begin{lstlisting}[language=bash]
python Evaluation/Evaluation.py
\end{lstlisting}
\end{itemize}

\textit{[Results]}
\begin{itemize}
  \item The evaluation script generates metrics (accuracy, precision, etc.).
  \item Expected results: Metrics should match those reported in Table~II of the paper.
\end{itemize}

\subsubsection{Experiment (E2): Correction Evaluation}

\textit{[Execution]}
\begin{itemize}
  \item Run the correction script:
\begin{lstlisting}[language=bash]
python Correction.py
\end{lstlisting}
  \item Input: Aligned functions from the detection phase.
  \item Output: Corrected functions saved to \texttt{Dataset\_4\_AE\_output/}, with \texttt{//fix} annotations.
\end{itemize}

\textit{[Results]}
\begin{itemize}
  \item Compare corrected outputs against \texttt{Ground truth/}.
  \item Evaluate using Fix Rate (FR) and Correct Fix Rate (CFR), following the definitions in Table~I of the paper.
  \item Note: Manual assessment is required for correction evaluation, as discussed in Section~V of the paper.
  \item Expected results: FR and CFR values should match those reported in Table~III of the paper.
\end{itemize} 






%



\end{document}